\documentclass[useAMS,usenatbib]{mn2e}
\usepackage{epsfig}
\newcommand\ocen{$\omega$~Cen~}

\newcommand{\sgnA}{{\it SG-R1~}}

\newcommand{\sgnD}{{\it SG-R05~}}
\newcommand{\sgnAA}{{\it SG-R1}}
\newcommand{\sgnBB}{{\it SG-R075}}
\newcommand{\sgnCC}{{\it SG-R06}}

\newcommand{\sgnco}{{\it SG-C10~}}

\newcommand{\sgnccoA}{{\it SG-C30}}

\newcommand{\popms}{\ensuremath{f_{\rm MS}}}

\newcommand\aj{{AJ}}%
\newcommand\araa{{ARA\&A}}%
\newcommand\apj{{ApJ}}%
\newcommand\apjl{{ApJ}}%
\newcommand\apjs{{ApJS}}%
\newcommand\aap{{A\&A}}%
\newcommand\mnras{{MNRAS}}%
\newcommand\pasj{{PASJ}}%
\newcommand\mi{{M_{\rm FG}}}
\newcommand\mii{{M_{\rm SG}}}
\newcommand\mn{{M_{\rm now}}}
\newcommand\mni{{M_{\rm FG,now}}}
\newcommand\mnii{{M_{\rm SG,now}}}
\newcommand\mg{{M_{\rm g}}}
\newcommand\msun{\rm{M_{\odot}}~}
\newcommand\mup{$M_{\rm up}$}
\newcommand\mmas{$M_{\rm mas}$}
\newcommand\teff{T$_{\rm eff}$}
\def\simgt{\lower.5ex\hbox{$\; \buildrel > \over \sim \;$}}
\def\simlt{\lower.5ex\hbox{$\; \buildrel < \over \sim \;$}}

\title[Multiple generations in globular clusters]{Formation and Dynamical Evolution of Multiple Stellar Generations in Globular Clusters}
\author[A. D'Ercole et al.]{Annibale
  D'Ercole$^{1}$\thanks{E-mail:annibale.dercole@oabo.inaf.it,$~$vesperin@physics.drexel.edu, \newline dantona@mporzio.astro.it,$~$steve@physics.drexel.edu,\newline simone.recchi@univie.ac.at}, Enrico Vesperini$^2$\footnotemark[1],
  Francesca D'Antona$^3$\footnotemark[1], \newauthor Stephen L.W. McMillan$^2$\footnotemark[1], Simone
  Recchi$^{4,5}$\footnotemark[1]\\  
$^{1}$INAF- Osservatorio Astronomico di Bologna, via Ranzani 1, I-40127 Bologna (Italy)\\
$^2$ Department of Physics, Drexel University, Philadelphia, PA, USA\\
$^{3}$ INAF- Osservatorio Astronomico di Roma, via di Frascati 33, I-00040 Monteporzio (Italy) \\ 
$^4$ INAF - Osservatorio Astronomico Trieste, Via G.B. Tiepolo 11, 34131,
Trieste (Italy)\\
$^5$ Institute of Astronomy, Vienna University, T\"urkenschanzstrasse 17,
A-1180, Vienna, (Austria)}
\begin{document}

\date{Accepted . Received .; in original form .}

\pagerange{\pageref{firstpage}--\pageref{lastpage}} \pubyear{2008}

\maketitle

\label{firstpage}

\begin{abstract}
We study the formation and dynamical evolution of clusters with
multiple stellar generations.
Observational studies have found that some globular clusters host a
population of second generation (SG) stars which show chemical
anomalies and must have formed from gas containing matter processed in
the envelopes of first generation (FG) cluster stars. We study the SG
formation process by means of 1D hydrodynamical simulations, starting
from a FG already in place and assuming that the SG is formed by the
gas ejected by the Asymptotic Giant Branch (AGB) stars.  This gas
collects in a cooling flow into the cluster core, where it forms SG
stars. The SG subsystem emerging from this process is initially
strongly concentrated in the cluster innermost regions and its
structural properties are largely independent of the FG initial
properties. We also present the results of a model in which pristine
gas contributes to the SG formation. In this model a very helium-rich
SG population and one with a moderate helium enrichment form; the
resulting SG bimodal helium distribution resembles that observed for
SG stars in NGC~2808.

By means of $N$-body simulations, we then study the two-population
cluster dynamical evolution and mass loss.  In our simulations, a
large fraction of FG stars are lost early in the cluster evolution due
to the expansion and stripping of the cluster outer layers resulting
from early mass loss associated with FG SN ejecta.  The SG population,
initially concentrated in the innermost cluster regions, is largely
unscathed by this early mass loss, and this early evolution leads to
values of the number ratio of SG to FG stars consistent with
observations.  We also demonstrate possible evolutionary routes
leading to the loss of most of the FG population, leaving an
SG-dominated cluster.  As the cluster evolves and the two populations
mix, the local ratio of SG to FG stars, initially a decreasing
function of radius, tends to a constant value in the inner parts of
the cluster.  Until mixing is complete, the radial profile of this
number ratio is characterized by a flat inner part and a declining
portion in the outer cluster regions.  
\end{abstract}


\section{Introduction} 
 \label{sec:introduction}
The common paradigm that Globular Clusters (GCs) are examples of
``Simple Stellar Populations'' (SSP), assemblies of stars born all at
the same time and sharing identical chemical composition, had until a
few years ago only one noticeable exception: $\omega$~Cen, the most massive
cluster, whose stars show a large spread in metallicity. Today, the
situation is much more complex, and it is now clear that practically
none of the GCs so far examined in detail is an SSP. Even when all the
stars in a single GC share the same Fe abundance, star-to-star
differences in the abundances of lighter elements are found. These
differences concern light-element abundances: in a relevant fraction
of cluster stars, they look like the result of nuclear processing
through proton capture reactions at high temperature (CN, ON, NeNa and
MgAl cycles). In particular, O-Na and Mg-Al anticorrelations are found
for stars at different stages of their evolution, from the red giant
branch to the main sequence (MS) \citep[e.g.][]{gratton-ar2004}.  Halo
field stars with the same metallicity do not show such huge variations
\citep{kra94}. This leads us to attribute the anomalous composition to
the fact that these stars formed in a dense GC environment, and that
nuclear processing in a first stellar generation (hereafter FG), has
been followed by a subsequent episode of star formation, creating a
second stellar generation (hereafter SG) from matter including this
processed gas.
 
 An additional ingredient to the complexity of the new scenario has
 been added by recognizing that there are at least two clusters,
 \ocen\ and NGC~2808, in which the MS morphology shows that a non
 negligible fraction of the stars ($\sim$25\% in \ocen, $\sim$15\% in
 NGC~2808) populate a ``blue MS'' which can only be interpreted as a
 very helium-rich MS \citep{bedin2004, norris2004, piotto2005,
   dantona2005}. The extreme helium population must also be very {\it
   homogeneous}, both in helium and metal content, since in both clusters
 the blue MS is well detached from the rest of the main sequence.
 \cite{piotto2007} have shown that the MS of NGC~2808 is actually made up of
 three different sequences: a normal MS, a population with somewhat
 enriched helium (Y$\sim$ 0.30) and a very helium-rich one
 (Y$\sim$0.35--0.40). In \ocen\ the ``normal'' MS is actually made
 up of sequences with different metal content, as also shown by the
 complexity of the subgiant branch(es) \citep[e.g.][]{villanova2007}.
 In addition to \ocen and NGC~2808, the two massive clusters NGC~6441
 and NGC~6388 apparently also harbor a very high helium population,
 including $\sim 10$\% and $\sim 20$\% of stars respectively.  In this case
 the evidence comes  from the high-\teff\  extension of the Horizontal
 Branch (HB)  
 \citep{caloi2007, busso2007,yoon2008}, a very peculiar occurrence at
 the high metallicity of these clusters \citep{rich1997}, and it is
 not clear whether the very high helium population is as homogeneous
 as it is in the previous cases.  

Observationally, it appears that these extreme Y sequences are present
only in the most massive clusters \citep{piotto2007}. A moderately
helium rich population ($0.26 < \rm{Y} < 0.30$), on the other hand, is
present in most GCs, largely independent of the total cluster mass.
The observational spectroscopic evidence resides mainly in the
existence of similar Na--O anticorrelation ---also among the unevolved
stars--- in all the clusters so far examined \citep[see, e.g., the
  summary by][]{carretta2006} although the most extreme Na--O
anomalies are present only in clusters that also show extreme
high-\teff\ extensions of the HB \citep[see][]{carretta2007}.  These
results, based generally on spectroscopy of a few stars, have been
recently confirmed by the analysis of larger samples of 30--100 stars
per cluster (Carretta et al., 2008)
  The indications coming from the analysis of the
CNO anomalies are similar \citep[for a summary, see
  e.g.][]{cohenetal2005}.

Photometrically, this non-extreme
 population can not be easily isolated in the MS photometry, mainly
 because the moderately helium rich isochrones merge with the standard
 isochrones in the turnoff region \citep{dantona2002,salaris2006}, but
 also because, in general, these SG stars do not have an unique value of
 Y \citep[e.g.][]{lee2005,caloi2007}.  Nevertheless, these stars show
 up quite prominently in the HB morphology, which
 amplifies the mass differences (due to the initial helium content)
 among stars of the same age \citep{dantona2002,daca04}.\footnote{The
age difference between FG and SG is in any case so small with
respect to the total GC age that it is irrelevant for the turnoff
evolution.}  It is important to notice that, according to the
detailed HB modelling by  \cite{caloi2007} \citep[see
also][]{dantonacaloi2008},  also in NGC~6441 and NGC~6388 the total 
SG population  includes $\sim 60 \%$ of the total number of stars, and 
is therefore much larger than just the very high helium fraction
population.
 
The interpretation of the HB morphologies in terms of helium
enrichment, coupled with the spectroscopic information indicating that
the helium-rich samples probably coincide with the sample of stars
showing the Na--O anomalies \citep[e.g.][]{carretta2006}, lead us to
conclude that, in the clusters so far examined, the percentage of
stars of the SG is generally $\sim$40--60\% \citep{dantonacaloi2008}.

For the clusters having mainly blue HBs, there are even indications
that they may contain only---or mainly---SG stars
\citep{dantonacaloi2008}. In an analysis of the relative magnitude
locations of the turnoff, red giant ``bump'' and HB,
\cite{caloi-dantona2005} have provocatively suggested this possibility
for M~13, the cluster showing the broadest range of chemical anomalies
\citep{sneden2004}.  In addition, some clusters in which the tightness
of the evolutionary sequences in the HR diagram \citep[e.g. the
  relatively small cluster NGC 6397,][]{king1998,richer2006} or the
short extent of the HB \citep[e.g. the massive GC M 53, ][]{dantonacaloi2008}
indicate that they indeed represent SSPs, might be dominated by a {\it
  homogeneous} SG. For example, in NGC~6397 only three out of fourteen
scarcely evolved stars examined by \cite{carretta2005} are normal in their nitrogen
content. In the others, [N/Fe]$\sim 1 -1.7$, showing CN and ON
processing in stars that should not show it. The blue HB of M~53 can
easily be described as uniparametric \citep{dantonacaloi2008}, but its
integrated spectrum shows a strong nitrogen enhancement with respect
to field halo stars \citep{liburstein2003}.  In conclusion, some
``normal'' blue--HB GCs (with typical current masses of a few $\times
10^5\msun$) should be considered self-enriched, and any model for the
formation and evolution of multiple population clusters must be able
to explain the observed large SG population and the cluster-to-cluster
variations in the SG/FG number ratio.
  
Both massive Asymptotic Giant Branch (AGB) stars
\citep{cottrell1981,ventura2001} or ``fast rotating'' massive stars
(FRMS) \citep{mm2006,decressin} have been proposed as possible
progenitors of the SG. The most critical point, common to both
scenarios, is the following: for a standard Initial Mass Function
(IMF) of the FG, such as a Salpeter or Kroupa IMF, the gas ejected by
the massive stars during the phase of H-burning, or the gas contained
in the envelopes of massive AGBs \citep{daca04, bekki-norris2006}, is
too scarce to form a large SG population.  Two solutions to this
conundrum have been suggested: either the system had a standard IMF
but it was initially much more massive---{\it at least a factor 10
  more massive}, under very conservative assumptions (see
Sect. 3.1)---than today's GC \citep{daca04, bekki-norris2006, prch06,
  dantona2007, decressin2007b}, or the system evolves at constant mass
and the IMF is highly anomalous \citep[][see also \cite{dosills07} for
  a dynamical study]{daca04,decressin2007b} and contains a significant
number of FG polluters. As we will discuss in Sect. \ref{subsec:imf},
a large initial mass is necessary also in this second scenario if a
cluster must have today an approximately equal number of long-lived SG
and FG stars (where we define long-lived stars as those having initial
masses in the range 0.1 $M_{\odot}$--0.8$M_{\odot}$).

We reconsider here the problem of GC formation to explain not only the
most peculiar very helium-rich population (15-20\% in three GCs, 25\%
in \ocen), but the whole SG population (about 40--60\% of the current
GC stellar content, and possibly even more in the ``blue HB''
clusters).

In this work, we follow the idea that the intermediate--mass stars of
the FG, starting at 8M$_{\odot}$ and extending at most down to
$\sim$4--5M$_{\odot}$, lose by stellar winds and the ejection of
planetary nebulae matter processed by the hot CNO cycle via Hot Bottom
Burning (Na- and Al-rich, O- and Mg-poor) during their AGB evolution,
and that an SG forms either from the pure AGB ejecta or from ejecta
mixed with pristine gas. While previous work examined the reliability
of this model from a chemical point of view, here we focus mainly on
its hydrodynamical and dynamical consistency.

In Sect. \ref{sec:sgprog} we first briefly summarize the state of the
art of chemical modelling of the SG, we discuss the reasons why we focus our
attention on the scenario in which the AGB stars are the
polluters and emphasize the possible role of  
super-AGB stars.

In Sect. \ref{sec:models} and
\ref{sec:res}, we address this  problem by means of 1D hydro
simulations.  We show that in the massive AGB epoch, the slow winds
collect in a cooling flow driving AGB ejecta to the center of the cluster,
where they form SG stars. Simulations exploring the dependence of our
results on the cluster mass, IMF, initial concentration as well as the
role of SNe Ia and other extra energy sources are also presented in
this section. 

In Sect. \ref{sec:pristine} we explore a model in which the pristine
matter is partially re-accreted after the SN II explosion stage.
The stellar population forming after the accretion shows a lower
helium abundance relative to the stars formed earlier from the
pure ejecta of the super-AGBs. This may explain the presence
of the very high-helium population observed in a few, very massive GCs.

In Sect. \ref{sec:dyn0} we present the results of a number of $N$-body
simulations exploring the subsequent cluster dynamical evolution.  We
show that the early cluster expansion driven by mass loss due to SN
ejecta leads to a strong preferential loss of FG stars.  Our
simulations show that dynamical evolution can lead to values of the
ratio of the number of SG to FG stars consistent with those suggested
by observational studies. We also show possible evolutionary routes
leading to the loss of most of the FG stars, leaving an SG-dominated,
cluster.

\section{The progenitors of the second generation: preamble}
\label{sec:sgprog}
 \subsection{Why AGB and super-AGB stars} 
\label{subsec:agb}
Although the study of the formation and dynamical evolution of
multiple population clusters is the main goal of this paper, one of
the key assumptions of our investigation is that the SG progenitors
are AGB stars, a scenario which has been criticized in a few recent
studies aimed at producing a plausible chemical evolution history of
GCs.  In this section we briefly review a number of recent studies of
the chemical modeling of SG stars and discuss why the AGB stars (along
with the implications of the role of super-AGBs) are still to be
considered a key ingredient in any theory for the formation of a SG
population.

In the last decade, attention has been focused mainly on the AGB
scenario, based on the simple idea that the AGB winds are slow enough
to be retained in the potential well of the cluster, and that these
winds are processed through the hot CNO cycle by ``Hot Bottom
Burning'' \citep[][]{cameron-fowler1971, sack1992} at the bottom of
the convective envelope \citep{ventura2001}. Many different groups
have addressed the computation of suitable models for the chemical
anomalies, and a number of results apparently do not support this
scenario \citep{fenner2004, karakas2006, choi2008}, while the model
presented in \cite{bekki2007} is in only partial agreement with the
data.

The model of SG formation by the contribution of FRMS has been
proposed much more recently \citep{mm2006,prch06,decressin,
  decressin2007b} and, while it is not our goal to dismiss the FRMS
scenario, we emphasize that many of its aspects are still to be fully
explored. In particular, the main assumptions of this model are
particularly stringent: all the stars must rotate at break-up speed,
both to achieve a deep rotational mixing of CNO processed matter in
the envelope and to inject, with small or zero ejection velocity, the
CNO processed ejecta into the cluster, where they mix with the
intracluster pristine matter \citep{decressin}.  Conversely, the
high velocity winds during the helium-core burning phase and the
supernova ejecta must leave the cluster without any interaction with
the remnant gas. The hydrodynamical and dynamical aspects of the FRMS
scenario have not been studied, and a recent VLT FLAMES survey of
massive stars, studying the nitrogen abundance and the rotational
velocity of 135 B-type stars in the LMC \citep{hunter2008}, has found
a number of unenriched fast rotators and cast some doubt on the
efficiency of rotational mixing.

As far as the AGB scenario is concerned, it is important to emphasize
the following points.  In discussing AGBs, we generally think of the
low-mass, thermally pulsing AGBs that go through long series of
thermal pulses and dredge-up episodes, so that they are generally rich
in carbon and s-process elements. But these are not the AGB stars that
can be progenitors of the SG in clusters.  Some studies show that the
sum of CNO elements is approximately constant among FG and SG stars
\citep{cohen2002, cohen2005}. This is possible only if we are dealing
with {\it the most massive AGBs}, in which the episodes of third
dredge-up are small in number, and the temperature at the base of the
convective envelope is so large ($>8\times 10^7$K) that the ON cycle is active,
reducing the O abundance as observed in the anomalous stars.
\cite{carretta2005} find that the CNO abundance is actually
somewhat---although not much---larger in the SG stars. The latter data
indicate that the third dredge-up \citep[e.g.][]{ibenrenzini1983}
plays a (small) role, but the dredge-up episodes must be limited in
number. Moreover, a few dredge-up episodes do not alter the s-process
abundances, so that the lack of s-process enrichment in the SG is not
an argument against the AGB polluters.\footnote{A different,
  borderline, situation seems to be present in NGC~1851, whose HR
  diagram shows a splitting of the subgiant branch \citep{milone2008},
  that can be interpreted as due to two different stellar populations
  differing in total CNO abundance \citep{cassisi2008}.  This is the
  only cluster in which a bimodal spread in the abundance of s-process
  elements has also been found, anticorrelated with the oxygen
  abundance \citep{yong2008}.  More than others, this case might
  indicate that the AGB scenario is to be preferred to the FRMS
  scenario: with respect to most GCs, the SG of NGC~1851 may have been
  born from somewhat less massive AGBs, for which the effect of the
  third dredge-up is more clearly evident, both in the total CNO
  abundance and s-process elements increase.}

A confusing issue when discussing the AGB evolution of low-metallicity
stars is that the chemical yields based on AGB modelling by different
groups differ enormously.  The yields based on sets of models computed
by Lattanzio's group, widely used for follow--ups and recently
published by \cite{karakas2007}, provide the correct abundance
signatures of the SG only for the most massive AGBs ($\sim 7\msun$),
while the yields of Ventura's group, and especially the most recent
models by \cite{ventura2008a, ventura2008b}, look compatible with the
anomalies down to $\sim 5 \msun$. The mass-loss rate formulation, the
nuclear reaction rates, and other details such as the treatment of
core-overshooting in the phases previous to the AGB, are all relevant
to the determination of yields \citep[e.g.][]{ventura2005b}.  However,
the most important difference among the physical inputs to these two
sets is the treatment of superadiabatic convection, which affects the
temperature stratification of models in the phase of ``Hot Bottom
Burning'' and the stellar luminosity and mass loss
\citep{ventura2005a}. A ``more efficient'' convection model ---like
the FST model \citep{cgm1996} adopted in Ventura's computations---
destroys oxygen more efficiently, shortens the AGB lifetime, and
reduces the number of thermal pulses and the effect of the third
dredge-up for a wider range of initial masses than in
\cite{karakas2007}.  Thus, the AGB yields are very model-dependent,
and the models of chemical evolution based on these yields are very
model-dependent too, so that negative results can not necessarily be
taken as falsification of the AGB progenitor scenario.

The first complete model for the chemical anomalies by
\cite{fenner2004} shows that self--enrichment by AGBs is at variance
with most of the observational anomalies, but this result could be
easily predicted, as it is based on the \citet{karakas2007} set.

More recently, attention has been mostly devoted to explaining the
presence of a population with extreme helium content, although, as we
have noted, this population is only the tip of the iceberg of the SG,
and it is present only in the most massive clusters.  As \ocen shows
the signatures of a complex chemical history \citep[see e.g.][for a
  summary of the extensive observational studies of
  \ocen]{romano2007}, and its formation took place most likely over an
extended period \citep[2--4 Gyr, e.g.][]{stanford2006}, several
attempts have been made to understand its very high helium population
by means of standard chemical evolution models. \cite{romano2007}
showed that the evolution of \ocen can not be understood in terms of a
``closed box'' model, and that, also after relaxing this hypothesis,
the helium-to-metal enrichment $\Delta Y/\Delta Z \sim 70$\ of the
blue MS stars of this system can not be attained, even with the very
high helium yields of the winds of massive stars. Models of
\ocen\ taking into account inhomogeneous pollution by SNe Ia can
reduce (by $\sim$40\%) but not eliminate the extra amount of helium
needed to explain the presence of the blue main sequence (bMS)
\citep{mar07}.

\cite{karakas2006} assumed that the helium yields come from the
massive AGBs, and that the IMF is indeed peaked at these masses, in
order to maximize the effect of the self--enrichment. In this case,
they indeed could find a fraction of SG stars with Y up to $\sim
0.35$, but this is accompanied by a very large CNO enrichment, that is
not observationally found. This result again depends on the models
adopted to describe the helium yield: only the \cite{karakas2007}
models provided helium abundances as large as Y$\simgt 0.35$ in the
AGB ejecta, and this value was partially achieved at the ``second
dredge-up'' \citep{ibenrenzini1983}, and partially by the third
dredge-up during the thermal pulse phase. The third dredge-up of
course also alters the total CNO abundance in the envelopes.

Only recently has it become clear that a very high helium content
(Y$\sim$0.38) can be achieved already  after the second dredge--up
in the envelopes of the most massive ``super-AGB'' stars
\citep{siess2007}. The sum of the CNO abundances can remain close to
the initial value during the following AGB thermal pulse phase, as the
efficiency of third dredge-up is limited, because the helium luminosity
during the thermal pulses is weak \citep{siess2007b}. 
Unfortunately, full models through the
super-AGB thermal pulse phase are not yet available, but it is clear
that, in principle, it is possible to achieve high helium abundances
without enhancing the CNO. Following this idea, \cite{pumo2008}
suggest that the extreme helium-rich stars are formed directly from
the ejecta of super-AGB stars.

Finally, \cite{choi2008}, even adopting a model that maximizes the
possible role of AGBs, tried and failed again to model the \ocen
$\Delta Y/\Delta Z \sim 70$\ for 30\% of its population, concluding
that ``alternative processes are desperately called for.'' Again they
do not include the most recent yields for super-AGBs in their
computations, and use an almost closed box model, assuming that for
the formation of the FG a fraction from 0.5 to 1 of the available gas
is used. Thus the gas from which the SG is formed comes at most from
half of the mass of the pristine gas plus the FG ejecta, or entirely
from the FG ejecta.  Under this hypothesis, for any reasonable shape
of the IMF, the helium rich AGB ejecta can not reach 30\% of the total
cluster mass, as we will discuss again in Sect. \ref{subsec:frmwrk}.
In fact, modelling must be able to predict {\it at least a factor ten
  mass loss from the FG}, to produce the desired number ratio of SG/FG
stars of $\sim 1$.

\subsection{The super-AGB phase}
 \label{sec:superagb}
In the exploratory models presented in this work, we do not consider
in detail the peculiarity of the super-AGB evolution, although we plan
to include it in much more detail in subsequent work dealing
explicitly with the chemistry of the ejecta.  However, we discuss
briefly here the super-AGB phase and its possible evolutionary
outcomes, as it has consequences for our interpretation of these
preliminary model results too.  The minimum mass evolving into a
core--collapse supernova is today referred to as \mmas, to distinguish
it from the limit \mup, representing the minimum mass for carbon
ignition, below which stars evolve into C--O white dwarfs (WDs).
Stars in the mass range \mup$<M<$\mmas\ are called super-AGBs, and
ignite carbon off-center in semi-degenerate conditions, but are not
able to ignite hydrostatic Neon burning in the resulting O--Ne
core. Consequently, degeneracy increases in the core, and these stars
may undergo thermal pulses \citep[e.g.][]{iben1997,ritossa1999,
  siess2006} and lose mass as ``normal'' (but quite massive and
luminous) AGBs, the difference being the core composition (O--Ne
versus C--O) and core mass ($\ge 1.05\; \msun$).

As mentioned in Sect. \ref{subsec:agb}, super-AGBs may play an
important role in the evolution of proto--GCs.  \cite{siess2007} shows
that their helium content in the envelope may reach the values Y=0.36
-- 0.38, reasonably consistent with the highest helium content found
in the extreme populations of a few GCs, and \cite{pumo2008} suggest
that they are in fact the progenitors of the extreme He populations.

The fate of super-AGBs depends on the competition between mass loss
and core growth \citep{poelarends2007, siess2007}: if mass loss wins,
they evolve into massive O--Ne white dwarfs \citep{nomoto1984}; if the
core reaches the Chandrasekhar mass, they evolve into electron capture
supernovae (ecSNe), electrons being captured on the Ne nuclei. Thus
also stars below \mmas\ may explode as supernovae, but these events
are at least a factor ten less energetic than SNe II \citep[$\le
  10^{50}$erg][]{dessart2006} and also a factor 10 smaller than SNe
Ia. In fact, it has been recently proposed that practically all the
neutron stars (NSs) today present in GCs are born from ecSNe: due to
their lower energy output, it is also probable that the newborn NS
receives a proportionally smaller natal kick, allowing it to remain
bound to the cluster \citep{ivanova2008}.
 
The presence of super-AGBs has an interesting consequence on the
cluster evolution: the fate of super-AGBs in fact is either
to end their life quietly as O--Ne WDs, or to explode as ecSN (either
directly, or due to accretion induced collapse induced by mass
transfer by a binary companion). In both cases, until O--Ne cores are
formed, we {\it do not} have SN Ia explosions, that are due to the
explosion of C--O WDs reaching the Chandrasekhar mass by
accretion. These events can take place only {\it after} C--O WDs are
formed in the cluster, when masses $M<$\mup\ evolve.  Consequently,
{\it the epoch during which the super-AGBs evolve is probably the
  quietest period of the cluster's life}, perturbed at most by ecSNe
explosions, either in single stars or by accretion induced collapse,
and in any case not so energetic to alter either the gas evolution, as
we will see in the hydrodynamical simulations, or its chemistry, as
the whole core remains locked into the remnant neutron star.  We refer
the reader to \cite{siess2007} for a careful study of \mmas\ as a
function of the stellar metallicity.  His value for the typical
metallicity of GCs showing multiple stellar populations is $\sim
10\;\msun$\ for models not including core overshooting. For models
with moderate core overshooting this value is reduced by $\sim
1\;M_{\odot}$, thus to 9 $\msun$. The value of \mup\ from the models by
\cite{ventura2008a, ventura2008b} is $\sim 6.3\;\msun$ evolving at
$\sim$65Myr (see Table 1 in \cite{ventura2008a}).  These latter models
provide yields consistent with the GC chemical anomalies down to the
evolution of 5 $\msun$, evolving at an age of $\sim 10^8$ yr.  In this
scheme, in summary, the SG may be formed by super-AGBs down to M$> 6.3
\msun$, and by AGBs down to $\simgt 5\msun$. In
Sect. \ref{subsec:expl}, we choose 40~Myr as the minimum age to start
SN Ia explosions, in order to maximize their possible role. This limit
shifts to $\simgt 65$Myr for the \cite{ventura2008a} models, allowing
a longer time for the formation of the SG.

\subsection{Post-AGB evolution} 
\label{sec:postagb} 
We now consider whether the post-AGB evolution can play a role in the
formation of the SG. In fact, after the loss of the envelope, the
remnant hot cores evolve toward the white dwarf stage, reaching very
high values of \teff\ and thus becoming powerful sources of UV photons,
acting as extra energy sources and potentially affecting the behaviour
of the AGB gas.  A study of the possible dynamical role played by
extra energy sources is presented in Sect. \ref{subsec:sour}.

Post-AGB evolution of the massive AGB core begins when the envelope
mass is reduced below a critical value that does not allow stationary
H--burning on the AGB, so that the residual envelope contracts (at the
constant luminosity dictated by the core mass itself) and the star
becomes increasingly hotter. Planetary nebula 
excitation occurs at \teff$>$30000K, when UV photons begins to be emitted.
The critical envelope mass is a strong function of the core mass.
In addition, the stellar luminosity is also increasing with the core
mass. 
Based on the computations by \cite{paczynski1971}, \cite{iben85}
showed that the ``fading time" that is required for the luminosity to
fall by a factor 10 after the star has reached \teff=30000K is
proportional to $M_{core}^{-10}$, so that, while it is $>$10000yr at
the core masses of typical planetary nebula nuclei (0.6$M_{\odot}$), it is
only a few hundred years for the core masses 1.0--1.3 $\msun$ that are
remnants of the possible progenitors of the SG we are
considering \citep[see also][for a review]{vanwinckel03}.
These objects, being so short-lived, are not a significant
extra energy source capable of affecting the gas dynamics during the
evolutionary phase we are considering.  
 
\section{Inputs to the gas dynamical models} 
 \label{sec:models}
\subsection{The framework}
\label{subsec:frmwrk}
The model-building is as follows. The present GC is assumed to be
composed of two stellar components, namely the FG and the SG
stars. Its mass can thus be written as $\mn=\mni+\mnii$, where $\mni$
and $\mnii$ represent the amount of mass today in long-lived FG and SG
stars, respectively (recall, as mentioned in
Sect. \ref{sec:introduction}, that we define long-lived stars as those
having initial masses in the range $0.1<M/M_{\odot}<0.8$). The value
of $x$ (in the range \newline $0\leq x\leq 1$) indicates the relative abundance
of the two populations: $M_{\rm FG,now}=x\mn$ and $M_{\rm
  SG,now}=(1-x)\mn$.

The FG is assumed to be already in place with an initial
mass $\mi$ following a continous King radial profile
\begin{equation} 
\label{eq:kng}
\rho_*= \rho_0 \left[ 1 + \left({r \over r_c} \right)^2 \right]^{-1.5}
\end{equation}
up to the truncation radius $r_{\rm t}$. These stars have masses in
the range $0.1<M<100$ M$_{\odot}$ distributed according to a Kroupa
IMF $\Phi(m)$ \citep[][]{kroupa93}. The mass fraction $\delta$ of
long-lived FG stars is given by
\begin{equation}
\label{eq:frc}
\delta=\int_{0.1}^{0.8}m\Phi(m)dm.
\end{equation}
For our assumed IMF, we find $\delta=0.5$. 

The evolutionary times are taken from \citet{pama93}.  We will
consider as progenitors of the SG the stars of masses below the limit
for core--collapse supernova (SN~II) explosion, here assumed to be
$M=8$ $\msun$.  We discussed in Sect. \ref{sec:superagb} that, for the
metallicities of GCs, this limit should probably be shifted to $\sim
9\;\msun$ for models including core overshooting, but the precise
choice does not make an important difference to the global picture.

We start our calculations at $t=28$ Myr, when all the SNe II have
exploded, clearing the GC of all the pristine gas (we will reconsider
the possible role of pristine gas in Sect. \ref{sec:pristine}).  At
this stage, $\mi$ is reduced by an amount $\Delta \mi=\epsilon\mi$;
for a Kroupa IMF $\epsilon=0.08$, but it can be much larger for
flatter IMFs (see Sect. \ref{subsec:imf}).  The SG stars form from the
ejecta of the AGB stars of the FG during a period of time $\Delta
t_{\rm f}$, which must be long enough to allow the formation of a
substantial SG population, but not so long as to
include evolution of stars of masses M$\le 4-5 M_{\odot}$, which do not
provide yields compatible with the observed chemical anomalies.  The
value adopted in our simulations, $\Delta t_{\rm f}=100$ Myr, satisfy
this constraint and only stars with initial masses $M > 4\msun$
contribute to the SG formation.

The amount of gas returned by the FG in this period is $\Delta M_{\rm
  g}=\beta\mi$, where, for Kroupa IMF, $\beta=0.05$. Following
\citet{prch06} and \citet{decressin}, we assume that SG stars have
masses within the range 0.1-0.8 $M_{\odot}$, and thus are all still
present today. In the working hypothesis that all the gas forms stars,
i.e. that the star formation efficiency is $\eta=1$, the SG mass is
therefore $M_{\rm SG,now}=\Delta M_{\rm g}=\beta\mi$.

We stress that our assumption of a SG IMF skewed toward low masses is
made to minimize the FG mass: if stars with $M>0.8$ $\msun$ are
assumed to also be formed, the required ejecta mass, and thus $\mi$,
would be larger. The assumption of a flatter IMF, with its larger mass
return, does not lead to a significant lowering of $\mi$. In fact, in
this case, the number of long-lived FG stars would decrease, and thus
a large value of $\mi$ is still needed to obtain a significant quantity
of them (see Sect. \ref{subsec:imf}). It is worth noticing that
while SG stars more massive than 0.8 $\msun$ can be included in our
model, stars with masses larger than 8 $\msun$ must be excluded; these
stars explode as SNe II and would sweep out the gas returned
by the FG stars, preventing the formation of a substantial stellar
SG. In any case, we point out that formation of (only) low mass stars
has been invoked also in the cooling flow occurring in elliptical
galaxies \citep[e.g.][]{mabr99} which are a scaled version of the
cooling flows taking place in our GC models (see Sect.
\ref{subsec:std}).

From all the above, it follows that $\mi=f\mn$, where
$f=(1-x)/\beta$.  For a Kroupa IMF, and assuming $x=0.5$ as a rule of
thumb, we obtain $\mi=10\mn$. Thus, a GC with a current mass $\mn$
originates from a FG ten times larger, or even larger if the gas is
not completely turned into stars (or if the IMF of the SG is not
limited to 0.8 $M_{\odot}$). This huge extra amount of mass must be lost
in the successive evolution of the cluster. We postpone the discussion
of this point to Sect. \ref{sec:dyn0}, and describe in the next sections the
dynamics of the gas returned by the FG and the formation of the SG 
stars.
 
\subsection{Physical inputs}
\label{subsec:model}

In order to follow the gas evolution in the GC, we adopt the 1D version
of the hydrocode described in \cite{bede86} and solve
numerically the following equations:
\begin{equation}
\label{eq:den}
{\partial \rho \over \partial t} +\nabla \cdot (\rho {\bf v})=\alpha \rho_*-\nu {\rho \over t_{\rm sf}} 
\end{equation}
\begin{equation}
\label{eq:imp}
{\partial {\bf m} \over \partial t} +\nabla \cdot ({\bf m}\otimes {\bf v})=
-(\gamma -1)\nabla E - {\bf g} \rho -\nu {{\bf m} \over t_{\rm sf}} 
\end{equation}
\begin{equation}
\label{eq:ene}
{\partial E \over \partial t} +\nabla \cdot (E {\bf v})=
-(\gamma -1)E\nabla \cdot {\bf v} -\nu {E \over t_{\rm sf}} - L + S
\end{equation}
In the above equations, ${\bf v}$, $\rho$, ${\bf m}$ and $E$
represent, respectively, the gas velocity and the densities of the gas
mass, momentum and internal energy. The gravitational acceleration
${\bf g}$ is due to both FG and SG stars, as well as to the gas. The
ratio of the specific heats is $\gamma = 5/3$. Given the assumptions
described above, the mass loss occurs only from FG stars at the
specific rate $\alpha = \dot{M}_{FG} /\mi$, and $\rho _*$ thus
represents the FG stellar density. The mass loss rate $\dot{M}_{FG}$
is computed following \citet{cioetal91}, adapted to the IMF chosen for
the present case.  The star formation rate (SFR) is characterized by
the timescale $t_{\rm sf}=\max(t_{\rm cool},t_{\rm ff})$, where
$t_{\rm cool}$ and $t_{\rm ff}$ are the cooling time and the free-fall
time of the gas, respectively; it is usually $t_{\rm cool}\ll t_{\rm
  ff}$ and thus $t_{\rm sf}\propto t_{\rm ff}\propto \rho^{-0.5}$. The
arbitrary parameter $\nu$ summarizes our uncertainties about star
formation, but we find that our results are essentially independent of
its value (the reason for this is discussed in the Appendix); as a
consequence, all the models presented in this paper have $\nu=1$. We
stress, however, that the star formation is allowed only in regions
where the gas is converging. For this reason we set
\begin{equation}
\label{eq:nu}
\nu = \left\{ \matrix{ 
0, & \nabla \cdot {\bf v} \ge 0 \cr 
1, & \nabla \cdot {\bf v} < 0 \cr} \right..
\end{equation}

Radiative losses are given
by $L=n^2\Lambda (T)$, where $n$ and $T$ are the number density and
the temperature of the gas, respectively, and $\Lambda(T)$ is the
cooling curve as given by \citet{robr95}. The source term $S$ takes
into account the heating of the gas due to the feedback of the
stars:
\begin{equation}
\label{eq:src}
S=0.5\alpha \rho_*\left (3\sigma^2+v^2+v_{\rm w}^2 + 2q \right ).
\end{equation}
\noindent
The first three terms in the above equation are due to the
thermalization of the kinetic energy of the AGB winds, where $v$ is
the gas velocity, $\sigma$
is the one-dimensional velocity dispersion of the FG stars, and
$v_{\rm w}=2\times 10^6$ cm s$^{-1}$ is the wind velocity of the AGB
stars \citep[e.g.][]{marsh04}.  We also computed a set of models in
which an {\it ad hoc} heating source $q$ of stellar origin is added
(see Sect. \ref{subsec:sour}).

Assuming that the SNe Ia belonging to the FG start to explode with a
delay of 80-100 Myr, we are allowed to neglect their effect in our
simulations, and Eq. \ref{eq:src} represents the total energy source
of the system. If, however, such a delay is shorter, the repercussions
of the stellar explosions must be taken into account (see Sect.
\ref{subsec:expl}).

We conclude this section by pointing out that in our hydrodynamical
simulations we have assumed a static stellar potential. As discussed
in Sect. \ref{sec:dyn0}, the cluster is expected to expand during its
early evolution in response to the early mass loss due to SN ejecta.
In a further study now underway, we will explore the role of the early
expansion of the cluster driven by primordial gas loss and SN II
ejecta on the formation and final structural properties of the SG
system.  We have carried out a number of preliminary simulations with
initial conditions as in the standard case, for a FG system expanding
homologously and in isolation; the results seem to indicate that the
central gas and SG stellar densities and, in general, the conclusions
concerning the SG properties for the standard case are not
significantly affected by the cluster expansion.
 
\subsection{The energy budget}
\label{subsec:ebudg}

The energy required per unit time to extract from the GC the gas returned
at rate $\alpha$ by the FG is \citep[see][]{cioetal91}
\begin{equation}
\label{equ:lgrv}
L_g=-\alpha(t) \int_0^{r_t}4\pi r^2\rho_{\rm FG}(r)\phi_{\rm FG}(r)dr,
\end{equation}
where $\rho_{\rm FG}(r)$ and $\phi_{\rm FG}(r)$ are the radial profiles
of the FG stellar density and gravitational potential, respectively.
The energy input linked to such mass return and due only to the
stellar velocity dispersion is
\begin{equation}
\label{equ:lsig}
L_{\sigma}={3\over 2}\alpha(t) \int_0^{r_t}4\pi r^2\rho_{\rm FG}(r)\sigma^2(r)dr.
\end{equation}
By definition, these two equations can be written as
\begin{equation}
\label{equ:lgrv1}
L_g=2\alpha(t)|U|,
\end{equation}
\begin{equation}
\label{equ:lsig1}
L_{\sigma}=\alpha(t)K,
\end{equation}
where $U$ and $K$ are the binding energy of the GC and the kinetic
energy of the stars, respectively. If the FG is in virial equilibrium,
then $L_{\rm g}=4L_{\sigma}$. This is a general result independent of
the amount of mass and of the shape of the FG potential well. The
first three terms in Eq. \ref{eq:src} are of the same order of
magnitude; therefore in principle an extra source $Q\ge L_{\sigma}$
would be enough to revert the cooling flow into a wind, thus halting
the SF. However, because of the large radiative losses occurring in a
cooling flow, a wind can start only for sources much larger than
$L_{\sigma}$ (see Sect. \ref{subsec:sour}). It must also be pointed
out that, for sufficiently small $\mi$, the energy of the AGB winds
may vent from the GC at least part of the gas returned by the stellar
FG (see Sect. \ref{subsec:lowm}).

\section{Results}
\label{sec:res}
\subsection{The standard model}
\label{subsec:std}
For our standard model we assume a FG stellar mass $\mi=10^7$ $\msun$
in order to describe a GC with a current mass $\mn=10^6$ $\msun$, as
discussed in Sect. \ref{subsec:frmwrk}. This initial stellar
population follows a King radial profile with the parameters below
(see Eq. \ref{eq:kng}): $\rho_0=10^3$ $\msun$ pc$^{-3}$,
$r_{\rm c}=6.3$ pc and $r_{\rm t}=200$ pc. The truncation radius of the model
coincides with its tidal radius at a distance of 4 kpc from the Galactic center.

\begin{figure}    
\centering{
\includegraphics[width=8cm]{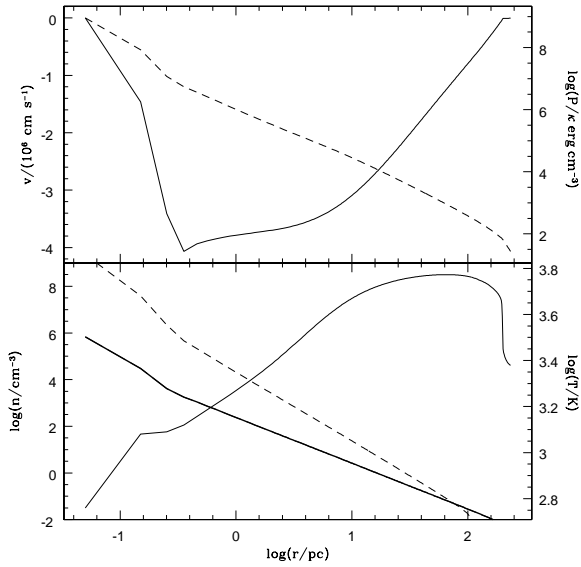}
}
\caption{ Radial profile of several quantities at $t=100$ Myr for the
  standard model . Upper panel: gas velocity (solid line) in units of
  $10^6$ cm s$^{-1}$ and pressure (dashed line) in units of $\kappa$
  erg cm$^{-3}$, where $\kappa$ is the Boltzmann's constant. Lower
  panel: gas density (thick solid line), gas temperature (thin solid
  line), SG stellar profile produced by the simulation (dashed
  line).  }
\label{fig:hyd} 
\end{figure}

Fig. \ref{fig:hyd} shows the radial profile of several quantities at
$t=100$ Myr, when the mass of the SG stars has grown to $M_{\rm
  SG}=4.45\times 10^5$ $\msun$. The upper panel shows that the fluid
velocity is negative over the entire cluster; radiative losses prevail
over the heating source, particularly in the central region where the
amount of gas shed by the FG is larger and causes significant
radiative cooling, which strongly increases with the gas density (see
Sect. \ref{subsec:model}). Consequently, the gas pressure falls and
gravity succeeds in driving the gas inward, establishing a cooling
flow from the start. This gas accumulates in the center, as shown by
the thick solid line in the lower panel of Fig. \ref{fig:hyd}.  The SG
stars forming from this gas are thus more concentrated than the FG
stars; their distribution is indicated by the dashed line, terminating
at radius $r\sim 100$ pc, half of the assumed $r_{\rm t}$.  The lower
panel also shows the temperature profile increasing with radius, as
expected in a cooling flow.

Since $t_{\rm sf}\propto \rho^{-0.5}$ (see Sect.  \ref{subsec:model}),
the SFR is proportional to $\rho^{1.5}$ (see Eq. \ref{eq:den}), and
the SG stars form preferentially in the central region, where they are
highly concentrated. The dashed line in the lower panel of
Fig. \ref{fig:hyd} indicates the radial distribution of the SG stars
as they form from the gas.

\begin{figure}    
\centering{
\includegraphics[width=8cm]{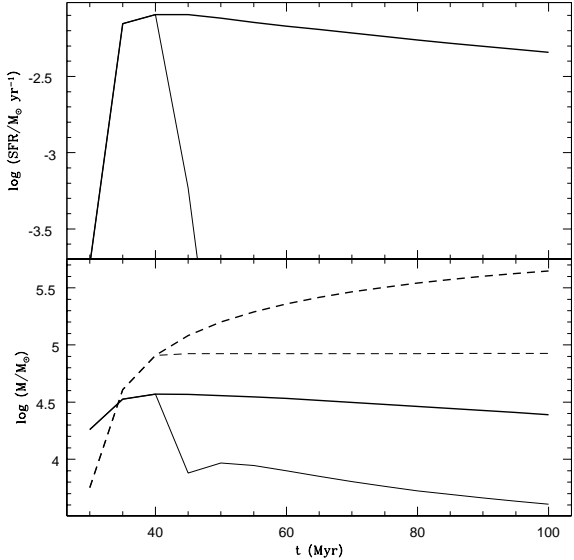}
}
\caption{Evolution of the standard model.
Upper panel: SFR evolution. Lower panel: temporal profile
of the amount of gas (solid line) and SG stars (dashed line).
The thick lines refer to an evolution in absence of SNe Ia.
In both panels the thin 
lines indicate the effect of SN explosions starting at $t=40$ Myr and
occurring every $5\times 10^4$ yr.
}
\label{fig:mt} 
\end{figure}

The thick lines in Fig. \ref{fig:mt} show the gas and the SG mass
evolution, as well as the time evolution of the SFR. Although the star
formation process is rather effective, we point out that $\mii=5\times
10^5$ $\msun$ (corresponding to the case $x=0.5$) is attained only at
$t=120$ Myr. Actually, the final value of the SG mass depends on the
time at which a suitable number of SNe Ia belonging to the FG start to
explode, removing gas from the cluster and inhibiting further star
formation.  We will discuss the role of SN Ia explosions further in
Sect. \ref{subsec:expl}.

\begin{figure}    
\centering{
\includegraphics[width=8cm]{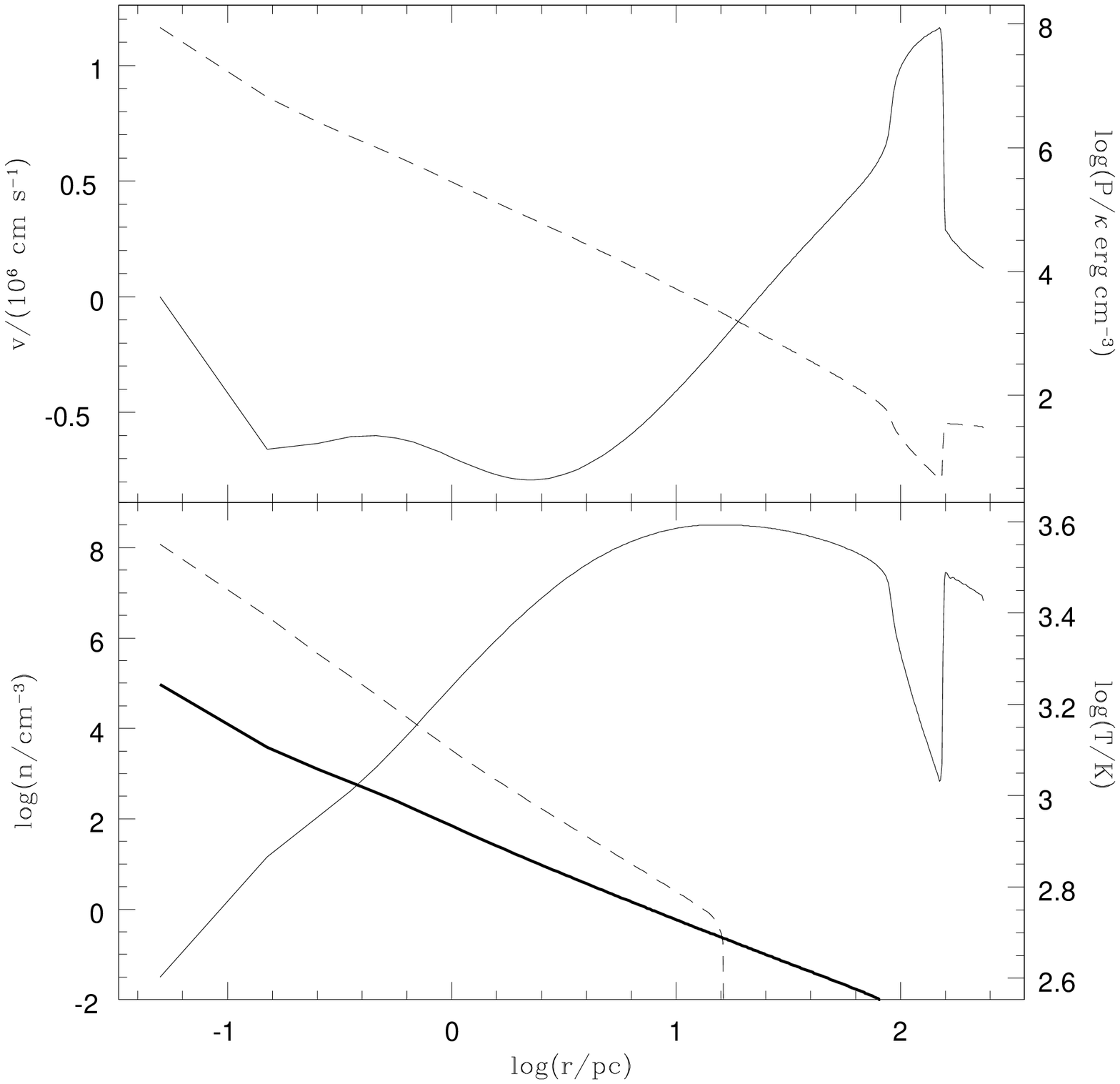}
}
\caption{ As Fig. \ref{fig:hyd}, but for $\mi=10^6$ $\msun$  }
\label{fig:hyd2} 
\end{figure}

\subsection{Model with a lower initial FG mass }
\label{subsec:lowm}

\begin{figure}    
\centering{
\includegraphics[width=8cm]{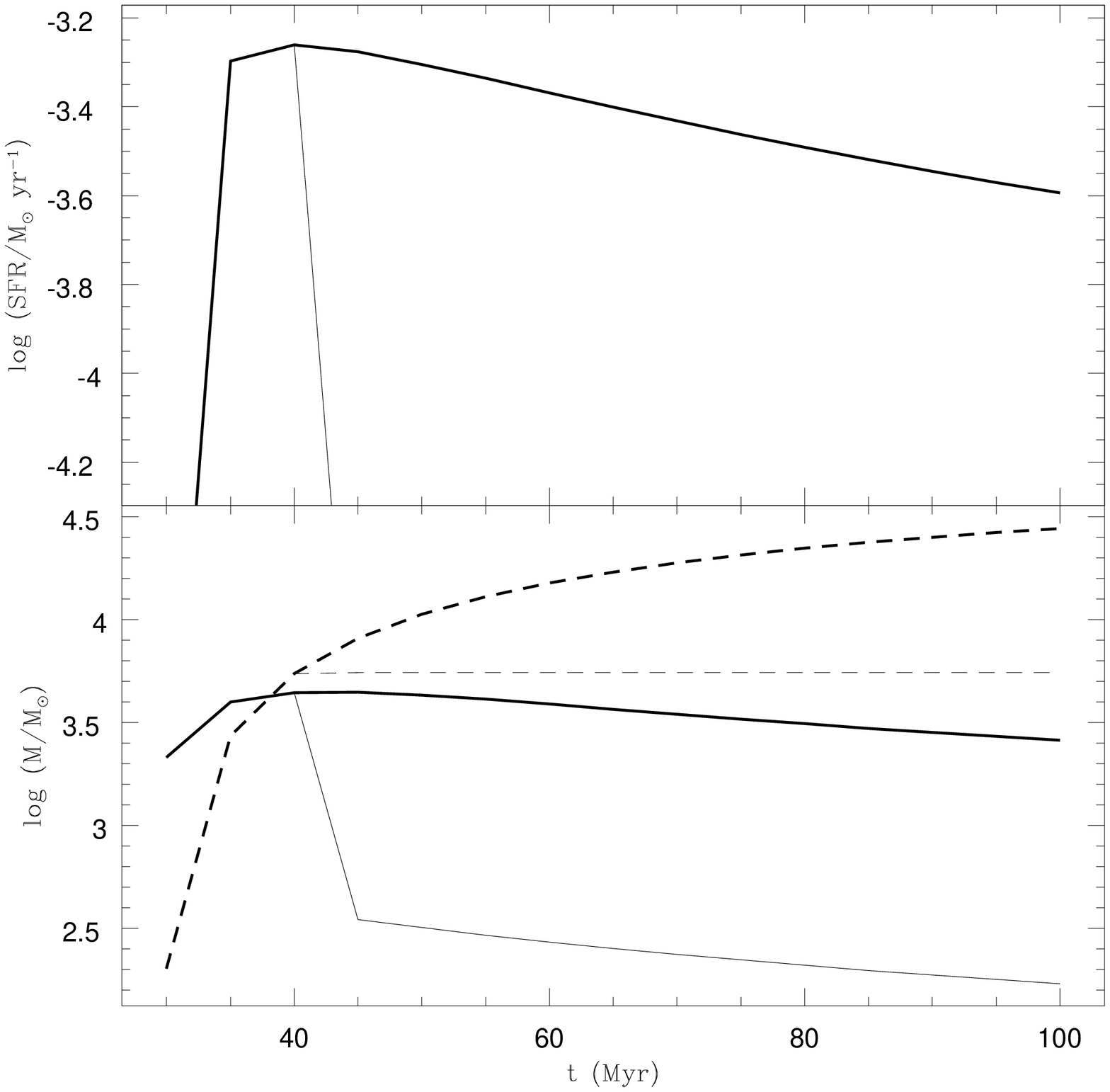}
}
\caption{
As Fig. \ref{fig:mt}, but for $\mi=10^6$ $\msun$.
}
\label{fig:mt5} 
\end{figure}

We have run a model similar to the standard one, but with a stellar FG
mass $\mi=10^6$ $M_{\odot}$. In this case, the initial parameters are
$\rho_0=1.08\times 10^3$ $\msun$ pc$^{-3}$, $r_{\rm c}=2.85$ pc and
$r_{\rm t}=90$ pc (also in this case the truncation radius corresponds
to the tidal radius at a distance of 4 kpc from the Galactic
center). In principle, since the gravitational potential, mass return
rate, and $L_{\sigma}$ all depend linearly on $\mi$, the gas behaviour
in this case should simply be a scaled version of the cooling flow
developed in the standard case, with lower velocities, temperatures
and densities.  However, the source term $S$ (see Eq. \ref{eq:src})
depends also on $v_{\rm w}$, the velocity of the AGB winds, which does
not scale with $\mi$. Moreover, radiative losses depend quadratically
on the gas density (see Sect. \ref{subsec:model}). For these reasons,
the self-similar scaling of the models with $\mi$ breaks when $\mi$ is
low enough to generate a stellar velocity dispersion $\sigma$ a factor
of a few lower than $v_{\rm w}$ and a mass return from the FG stars
such that the gas density is not so large to provide effective
radiative cooling. In the present model the source term prevails on
the radiative losses only at the outskirts of the cluster; as a
consequence, a partial wind forms, with the stagnation point located
at $r=23$ pc (see Fig. \ref{fig:hyd2}). For this reason the SG stellar
population is relatively more concentrated than in the standard case,
extending only out to $r=0.18r_{\rm t}$. However, the comparison
between Fig. \ref{fig:mt} and Fig. \ref{fig:mt5} (thick lines) shows
that the global behaviour of gas and SG stars are similar in the two
models.

\subsection{Model with a flatter IMF}
\label{subsec:imf}
\begin{figure}
\centering{
\includegraphics[width=8cm]{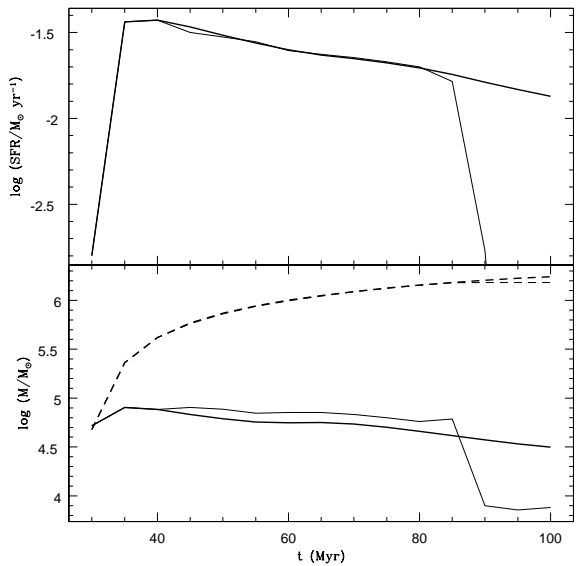}
}
\caption{
As Fig. \ref{fig:mt}, but for a flatter IMF (see text).
}
\label{fig:mt6f} 
\end{figure}

Following \citet{daca04}, one can assume a flatter IMF for the FG in
order to get a larger number of giant stars at early epochs. In this
way, the larger amount of gas returned by such stars would facilitate
the formation of the SG. To explore this scenario, we assume that for
$M<5$ $\msun$ the IMF follows a power law with an exponent $s=-1$,
while more massive stars follow a Salpeter distribution with
$s=-2.35$.  The flatter IMF leads to a larger amount of $\mg$ and
$\mii$ at any time compared to the standard model, as illustrated in
Fig. \ref{fig:mt6f} (thick lines).

Repeating for this model the simple analytical calculation of Sect.
\ref{subsec:frmwrk}, we obtain $\beta=0.2$, $\delta=0.05$ and
$\epsilon=0.5$. Assuming $x=0.5$, i.e. that half of the MS stars with
$0.1<M/M_{\odot}<0.8$ currently in the GC belong to the FG, we have
$0.05\mi=0.5\mn$ and thus $\mi=10\mn$. As $\Delta M_{\rm
  g}=0.2\mi=2\mn$, the star formation efficiency must be
$\eta=0.25$. In conclusion, also in this case, if one requires the
current mass of long-lived SG and FG stars to be similar, the total FG
initial mass must be ten times larger than the current total mass of
long-lived stars.  However, in this case the cluster must evolve
without losing mass because the initial number of long-lived FG stars
is much smaller.  We also point out an important difference concerning
the relative abundance of stellar remnants (white dwarfs, neutron
stars and black holes) and long-lived MS stars
($0.1<M/M_{\odot}<0.8$). Specifically, in both cases at $t \sim 13.5$
Gyr, $M_{\rm rem} \sim 0.15\mi$, but for systems with a flatter IMF
the total mass of long-lived stars is smaller and the ratio of the
total mass in stellar remnants to the total mass in long-lived stars
is much larger (by a factor $\sim 10^2$) for systems with a flatter
IMF.

\subsection{Models with SN explosions }
\label{subsec:expl}
In order to understand the effect of SNe Ia in our standard model, we
simulated the expansion of SN remnants (SNRs) through the GC. Given
the spherical symmetry of the model, we assume that the explosions
occur in the center; such an assumption is reasonable, as in the
central region there is a larger density of SN Ia progenitors
belonging to the FG. We also assume that the first explosion takes
place at the time $t_{\rm pk}$ at which the SN Ia rate quickly reaches
its maximum. This time is not unambiguously determined, and its
value for a SF of short duration is estimated in the range $40<t_{\rm
  pk}<300$ Myr \citep{mare01,mapa06}. In order to emphasize the
effects of the SNe Ia, in the present models we adopt $t_{\rm pk}=40$
Myr, following the SN Ia rate illustrated by \citet{mar06}.

We first consider the explosion of a single SN Ia taking as initial
values of the variables those obtained in the standard case at $t=40$
Myr. The occurrence of the supernova is mimicked by adding an extra
amount $E_{\rm SN}=10^{51}$ erg of thermal energy in the first three
mesh points of the grid.

We point out that the binding energy of the gas at the moment of the
explosion is only 44\% of that of the SN, so in principle the GC could
be completely deprived of gas by the stellar explosion. However, given
the large density of the gas itself, the radiative losses are rather
effective, and in fact no gas is lost by the GC.

\begin{figure}    
\centering{
\includegraphics[width=8cm]{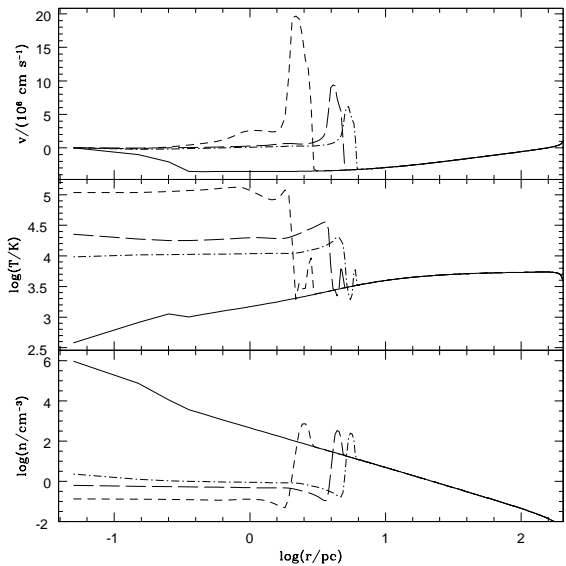}
}
\caption{ Velocity, temperature and density of the gas in the GC at
  $\Delta t= 10^4$ yr (dashed line), $\Delta t=3\times 10^4$ yr
  (long-dashed line) and $\Delta t=5\times 10^4$ yr (dot-dashed line)
  after a SN Ia explosion occurred in the standard model at $t=40$
  Myr. The solid lines show the radial profiles at $t=40$ Myr just
  before the SN explosion.  After $\Delta t=5\times 10^4$ yr another
  SN Ia is expected to explode (see text). }
\label{fig:expl} 
\end{figure}

Fig. \ref{fig:expl} shows the SNR expansion at three different
times. As expected in a Sedov-like solution, a ``hole'' is present
behind the shock front. Contrary to a genuine Sedov solution, however,
this hole is partially replenished by the FG high mass return rate
that at this stage is nearly $8.9\times 10^{-3}$ $\msun\,{\rm
  yr}^{-1}$.  The gas replenishment favours the radiative losses, thus
weakening the SNR shock and slowing the remnant expansion below the
escape velocity.

Although the effect of a single SN Ia explosion may therefore appear
as a minor perturbation of the cooling flow, the cumulative effect of
numerous SNe Ia drastically alters the gas dynamics.  At $t=40$ Myr
the SN Ia rate is of the order of $2\times 10^{-5}$ yr$^{-1}$ for
$\mi=10^7$ $\msun$ \citep{mar06}, i.e. one SN explosion every $5\times
10^4$ yr (the maximum time span considered in
Fig. \ref{fig:expl}). More recent SNRs overlap older ones before these
latter merge with the medium, and the sum of their actions has
substantial consequences on the GC evolution. As shown in
Fig. \ref{fig:mt}, the SFR quickly declines and the SG mass stops to
grow attaining the value $\mii=8.4\times 10^4$ $\rm{M_{\odot}}$. This
result is obtained assuming a constant SN Ia rate up to 100 Myr; this
assumption is largely justified as the actual rate suffers only a
negligible reduction during this interval of time.

We point out that such a dramatic reduction of the SFR is due not only
to a decrease in the gas content (the SFR being proportional to
$\rho^{1.5}$), but also to the condition for star formation we have
adopted (see Eq. \ref{eq:nu}). In fact, the gas is set into expansion
over much of the GC volume by the SN explosions, and dynamical
conditions are not favourable to further star formation.

In the end, as $\mii$ is only one-fifth of the value attained in the
standard case, in order to obtain a cluster  with a current total mass $\mn \sim
1.7\times 10^5$ $\msun$ with an equal amount of FG and SG stellar
components, an initial FG population with $\mi=10^7$ $\msun$ must
lose $\sim 99$\% of its mass.

\begin{figure}
\centering{
\includegraphics[width=8cm]{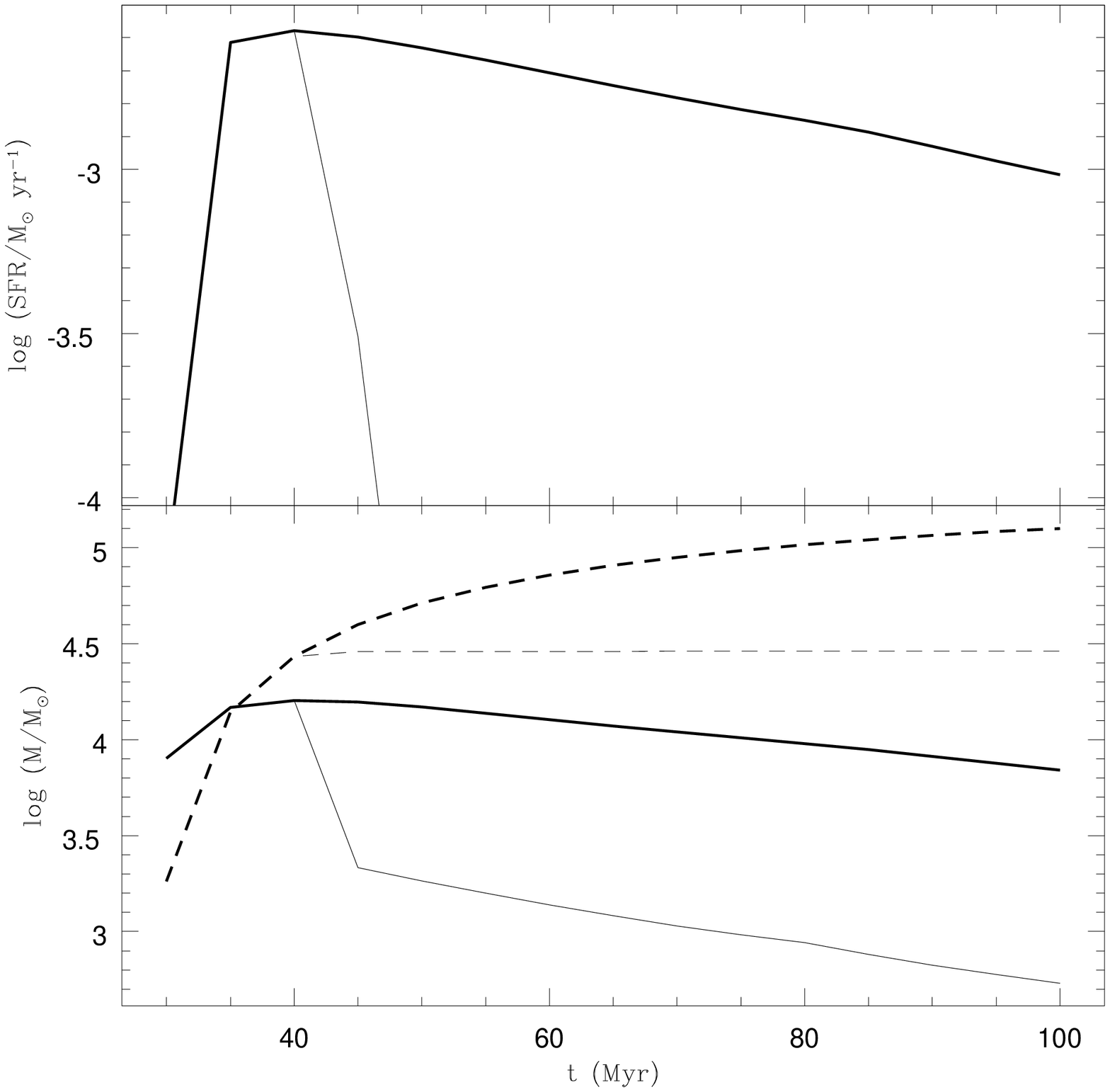}
}
\caption{
As Fig. \ref{fig:mt}, but for $\mi=10^6$ $\msun$ and a flatter IMF (see text).
}
\label{fig:mt5f} 
\end{figure}

It is interesting to note how differences in $\mi$ and/or the slope of
the IMF influence the effectiveness of SNe in halting the formation of
the SG stars.  Fig. \ref{fig:mt6f} shows that, contrary to the
standard case, the flatter IMF essentially neutralizes the effect of
the SNe Ia, as $\mii$ is only very slightly perturbed by the stellar
explosions. In fact, when the SNe Ia start to explode the amount of
gas is $\mg=7.6\times 10^4$ $M_{\odot}$, a factor of 2.1 larger than in
the standard case; as a consequence the radiative losses are roughly
$2.1^2=4.2$ times greater, and suppress rather effectively any
disruptive effect of the SNe.

Models with $\mi=10^6$ $\msun$ show similar behaviour (see Fig.
\ref{fig:mt5} and Fig. \ref{fig:mt5f}), independent of the IMF
adopted. In this case the gas has time to recover, at least in part,
its inward convergent flow between successive SN explosions, because
the SN Ia rate is ten times lower than in the models with $\mi=10^7$
$\msun$. Despite their lower rate, however, the stellar explosions are
equally effective in reducing the gas in the cluster because of its
lower binding energy in this case.

In principle, models in which the SF proceeds also during the SN
activity would show today some stars polluted by the SN Ia ejecta.
For example, let us examine the model with $\mi=10^7$ $\msun$ and
flatter IMF, which is the most favourable case to lock SN ejecta in SG
stars, as illustrated in Fig. \ref{fig:mt6f}. The mass return rate at
40 Myr is $3.4\times 10^{-2}$ $\msun$ yr$^{-1}$, while the mean
production rate of metals by SNe Ia is $2.8\times 10^{-5}$ $\msun$
yr$^{-1}$; as a consequence, the gas accumulating within the GC has a
metallicity increment due to the SNe Ia $\Delta Z=8.2\times 10^{-4}$,
which is comparable to the typical metallicity of GCs showing
self-enrichment.  On the other hand, star to star iron variations in
GCs do not exceed $\sim 0.04$ dex \citep{gratton-ar2004}, so that
retention of SN Ia ejecta in SGs is ruled out by observations, with
the exception of \ocen \citep[e.g.][]{mar07}.

\subsection{Models with an extra energy source}
\label{subsec:sour}
\begin{figure}    
\centering{
\includegraphics[width=8cm]{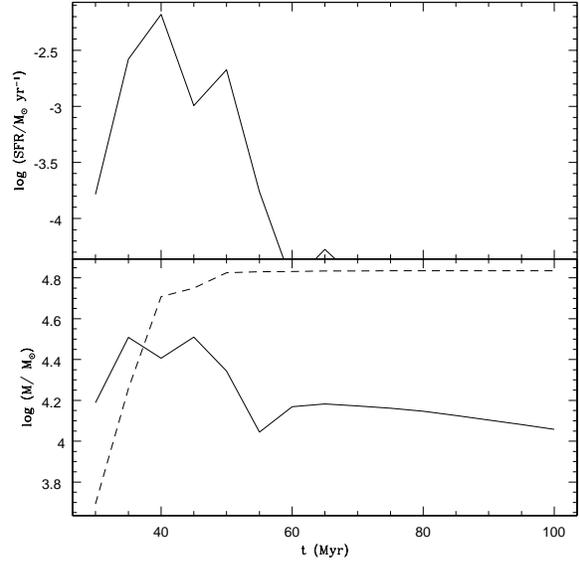}
}
\caption{As Fig. \ref{fig:mt}, but with a diffuse energy source
  $Q=6\times 10^{37}$ erg~s$^{-1}$ included (see text). }
\label{fig:lum} 
\end{figure}

In this section we explore the effect of an extra energy source
due, for example, to X-ray binaries or a number of luminous hot planetary
nebulae nuclei. In our model, this source is 
distributed as the FG stellar density and its total amount is equal to

\begin{equation}
\label{equ:extra}
Q=\int_0^{r_{\rm t}}4\pi r^2 \rho_{\rm FG}(r)q \hbox{d}r
\end{equation}

We have carried out a number of exploratory simulations and found
that, in general, a critical    
value $Q_{\rm cr}$ exists, separating models hosting a cooling flow
from models hosting a wind. The transition between the two kinds of
solutions happens rather abruptly as $Q$ increases from values
$Q<Q_{\rm cr}$ to values $Q>Q_{\rm cr}$. Such a sudden transition is
due to the strong nonlinearity of the radiative cooling term, which
depends on the square of the gas density.

Fig. \ref{fig:lum} is analogous to Fig. \ref{fig:mt} for a model
similar to the standard one, but with $Q=Q_{\rm cr}=6\times 10^{37}$
erg s$^{-1}$. After an initial transient phase during which some SF
occurs, a steady wind is established which prevents any further
substantial growth of $\mii$. This case is borderline between a model
with $Q=4\times 10^{37}$ erg s$^{-1}$, in which an almost normal star
formation occurs, and one with $Q=1.3 \times 10^{38}$ erg s$^{-1}$,
where the immediate onset of a fast wind precludes any SG formation.
For this set of models, the role of SNe Ia is negligible, as they
start to explode when the SG star formation has already been halted.
We stress that, for the standard model, we have $L_{\sigma}=2\times
10^{36}$ erg s$^{-1}$ at $t=28$ Myr, much lower than $Q_{\rm cr}$, as
expected owing to the radiative losses (see Sect.
\ref{subsec:ebudg}).

We mention that $Q_{\rm cr}=2\times 10^{36}$ erg s$^{-1}$ for
the model with $\mi=10^6$ $\msun$, while, for the models with
flatter IMF, we have $Q_{\rm cr}=6\times 10^{38}$ erg s$^{-1}$
and $Q_{\rm cr}=2\times 10^{37}$ erg s$^{-1}$ for $\mi=10^7$
$\msun$ and $\mi=10^6$ $\msun$, respectively. As expected,
$Q_{\rm cr}$ is larger for flatter IMFs, because of the larger
mass return rate which increases the radiative losses.

\subsection{Models with different initial concentrations}
\label{sec:core}

In our standard model the FG is distributed following a King profile
with concentration $c=\log(r_{\rm t}/r_{\rm c})=1.5$ (see Sect.
\ref{sec:models}). We analyzed also cases with the same values of $\mi$ and $r_{\rm
  t}$ but different concentrations, namely $c=0.75$, $c=1$, $c=2$ and
$c=2.25$. For these latter models, at $t=28$ Myr one has
$L_{\sigma}=0.86\times 10^{36}$ erg s$^{-1}$, $L_{\sigma}=1.1\times
10^{36}$ erg s$^{-1}$, $L_{\sigma}=3.5\times 10^{36}$ erg s$^{-1}$ and
$L_{\sigma}=5\times 10^{36}$ erg s$^{-1}$, respectively.

\begin{figure} 
\centering{
\includegraphics[width=8cm]{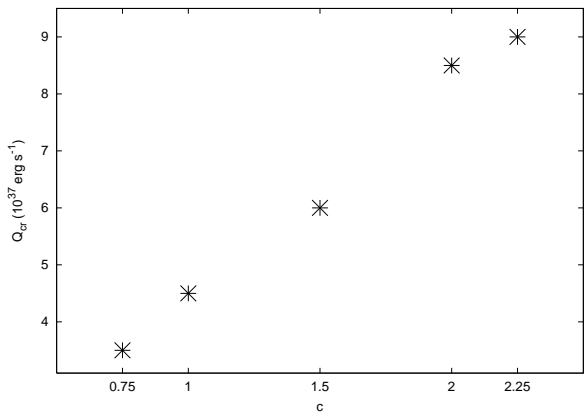}
}
\caption{$Q_{\rm cr}$ as a function of the cluster concentration $c$
for models with the same mass, IMF and truncation radius as the standard
model. }
\label{fig:qc} 
\end{figure}

Within the range of values of $c$ we have considered, the results are
remarkably similar to those of the standard case, and in fact $Q_{\rm
  cr}$ varies only by a factor of $\sim 2$, as shown in
Fig. \ref{fig:qc}. There is a trend of increasing $Q_{\rm cr}$ with
$c$ due to the fact that the stellar density gets larger in the GC
core; as a consequence, the gas density increases because of the
larger mass return, and the radiative losses are stronger. A
partial wind establishes in the outer regions, while a cooling flow
persists in the core. Reverting such a 'mini-inflow' becomes increasingly
more difficult for larger values of $c$. We point out that, although
$L_{\sigma}$ increases faster than the critical source, it remains
always much lower than $Q_{\rm cr}$.

\subsection{Models with different initial truncation radii}
\begin{figure} 
\centering{
\includegraphics[width=8cm]{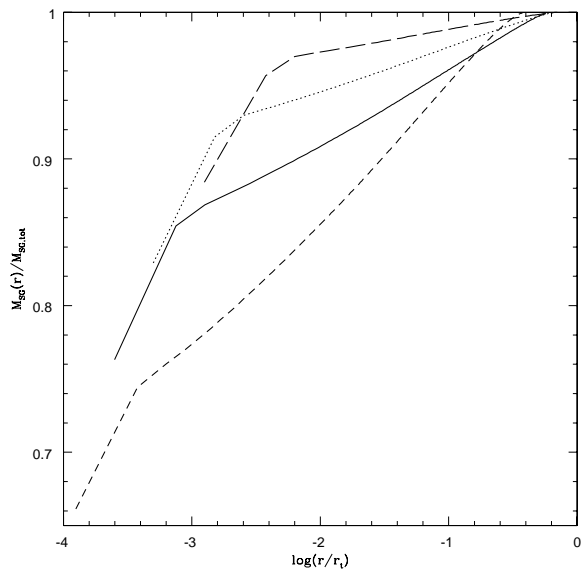}
}
\caption{ Radial profile of the SG stars at $t=100$ Myr for models
  with $\mi=10^7$ $\msun$, $c=1.5$ and $r_{\rm t}= 40$ pc (long-dashed
  line), $r_{\rm t}= 100$ pc (dotted line), $r_{\rm t}= 200$ pc (solid
  line, standard model), $r_{\rm t}= 400$ pc (short-dashed line).  The
  masses are normalized to their respective maximum values:
  $\mii=4.8\times 10^5$ $\msun$ (long-dashed line), $\mii=4.8\times
  10^5$ $\msun$ (dotted line), $\mii=4.45\times 10^5$ $\msun$ (solid
  line, standard model), $\mii=3.52\times 10^5$ $\msun$ (short-dashed
  line).  }
\label{fig:compm} 
\end{figure}

In order to understand the influence of the assumed size of the FG
population on the relative concentration of the SG stars, we ran
models with the same mass $\mi$ and concentration $c$ of the standard
model, but with different values of the truncation radius
($r_t=40,~100,~400$ pc).  The comparison among the models is
summarized in Fig. \ref{fig:compm} and shows only a minor dependence
of the relative concentration of the two populations on the initial
size of the cluster.  In all cases the SG population is strongly
concentrated in the cluster innermost regions, with 80--90 percent of
the total SG mass enclosed within $r/r_t \sim 10^{-3}-10^{-2}$.

\subsection{Consequences of the end of the cooling flow}
\label{sec:conclusion1}
The energy input either by recurrent SN~Ia explosions or by
distributed stationary sources with total power $>Q_{\rm cr}$
drastically limits the epoch of gas accretion and SG star formation in
the cluster core. Clearly, there are a number of uncertain parameters
in our understanding of SN~Ia rates, and also their starting age and
actual rate at early epochs may be smaller than we suggest (see
Sect. \ref{sec:superagb}).

Our simulations show that SN~Ia explosions effectively clean clusters
with initial masses in the range $10^6-10^7$ $M_{\odot}$ from AGB
ejecta , and limit the SG formation time.  We point out
that this is no longer the case for more massive objects like the
ultra-compact dwarf galaxies, and possibly the progenitor of \ocen,
which are embedded in massive dark-matter haloes, and which proceed to
a more complex (but also more standard) chemical evolution
\citep[see][]{mar06,mar07}.

Concerning the stationary sources, $Q_{\rm cr}\sim 10^{38}$
erg~s$^{-1}$ is of the order of the Eddington accretion luminosity for
NSs, so a number of sub--Eddington accreting NSs distributed in the
cluster could succeed in stopping the SG formation epoch. We know that
several hundreds of NSs, revealed by their millisecond pulsar radio
emission, are present in today's clusters \citep[e.g.][]{heinke2005}.
Their radio emission (relevant for those NSs that are already spun up
to millisecond periods at the early epoch we are considering) is in
the plausible range $10^{34} - 10^{35}$erg~s$^{-1}$, so it has no
global effect on the cooling flow. Only accreting NSs are good
candidates as stationary source. Although only a few stationary X--ray
sources are currently active in all the galactic GCs, the situation
might have been different in the early stages of cluster evolution and
this issue will need further investigation. As discussed in
Sect. \ref{sec:superagb}, NSs are just born from the highest mass
super-AGBs, when the super-AGB winds are collecting in the core.

From the chemical point of view, the fact that the cooling flow is
halted is important since it prevents less massive AGB stars from
contributing to the formation of the SG. Although the maximum allowed
age (and minimum allowed mass) strongly depends on the efficiency of
convection, third dredge-up and mass loss in AGB stars, if the cooling
flow is not halted and star formation prolongs beyond $\sim10^8$ yr
(the upper limit adopted in the standard hydro models of
Sect. \ref{subsec:std}), in all the current stellar models the third
dredge-up is important, and the yields are not compatible with the
observed chemical anomalies; specifically, the total CNO abundances
and the sodium abundance will be much larger than observed, and the
oxygen abundance will not be reduced.

\section{Pristine gas}
\label{sec:pristine}

\subsection{Motivations}
If the SG formation time is too short, it will not meet the
observational requirements, as not enough matter will be contained in
the ejecta.  One possible solution is the following: while the SN II
events have wiped out the pristine matter from the cluster center,
during the AGB epoch it is possible that the pristine gas in the
outskirts participates in the cooling flow.  In the central regions it
will mix with the AGB gas and contributes to star formation.  In this
way, the span of SF time required to achieve a massive SG is shorter.
Further, the ``dilution'' of the AGB ejecta by pristine gas may fully
explain, in particular, the milder chemical anomalies \citep{prch06,
  decressin, decressin2007b, ventura2008a, ventura2008b}.

In all the models discussed  in the previous sections, it has been
assumed that the FG SNe II  
deprive the GC of all its pristine gas. In this section we drop this
assumption and study a model  in which part of the pristine gas
contributes to the SG formation. We assume that
initial asymmetries in the distribution
of gas allow to vent out the SN II ejecta along preferential
directions, creating an ``hourglass'' cavity, and leaving part of the
pristine gas in a substantially unperturbed torus at the outskirts of
the cluster \citep[e.g.][]{rec01}. 

\subsection{Model with pristine gas infall}
\label{subsec:cavity}

Since we use a 1-D code for our hydrodynamical simulations, we can not
properly explore the case of SN II ejecta blown away through an
axisymmetric chimney, with part of the pristine gas confined around
it. We approximate this situation assuming an initial gas distribution
within the GC with a central hole of radius $r=150$ pc and with a
density radial profile $\rho(r)\propto r^{-2}$ beyond it. The total
gas mass within the numerical grid is $M=2.6\times 10^5$ M$_{\odot}$.
These values are rather arbitrary, but nevertheless the model is
useful to understand the consequences of the primordial gas infall on
the SG formation process.  All the other details of the model are the
same as in the standard model.

\begin{figure}    
\centering{
\includegraphics[width=8cm]{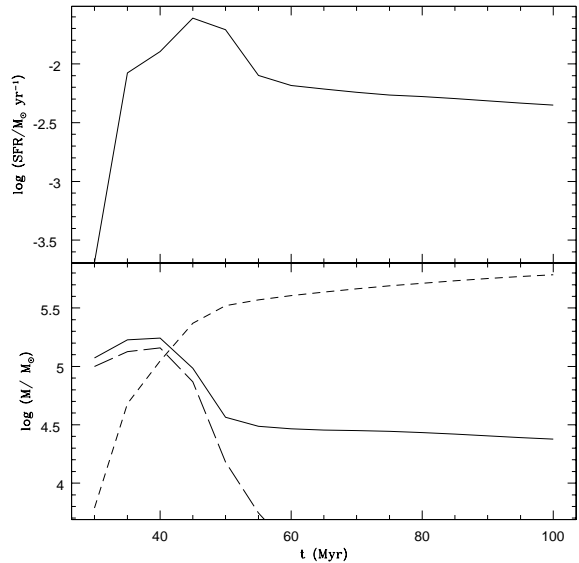}
}
\caption{ As Fig. \ref{fig:mt}, but for the infall model. The long-dashed
line represents the amount of pristine gas.  }
\label{fig:cav} 
\end{figure}

As apparent in Fig. \ref{fig:cav}, after 100 Myr the total mass of SG
stars formed, $\mii=6.2\times 10^5$ M$_{\odot}$, is larger than that
of the SG population formed in the standard model. This larger amount
is created very quickly because part of the pristine gas is in place
from the beginning and available for star formation. As this gas moves
toward the center, and more gas enters the GC from outside, the SFR
increases. After $\sim 50$ Myr no more pristine gas falls within the
GC, and that already present has been locked within the SG stars
formed till then (see the dashed line in the bottom panel of
Fig. \ref{fig:cav}). As a consequence, the global evolution of the GC
recovers its ``standard'' behaviour (see Fig. \ref{fig:mt}).

\begin{figure}    
\centering{
\includegraphics[width=8cm]{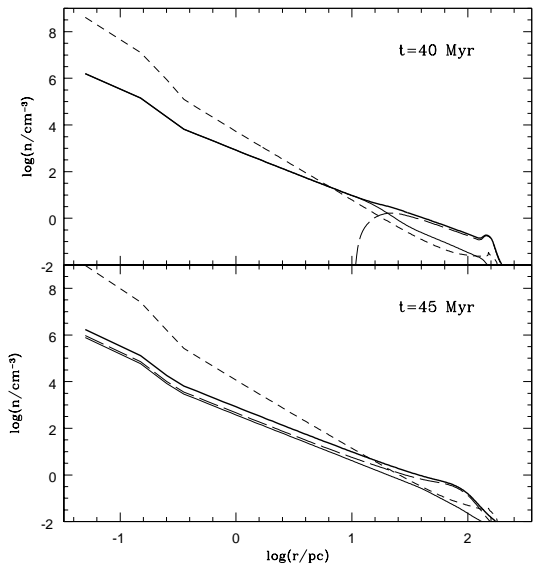}
}
\caption{ Radial profile of several quantities of the infall model at two
different times: total gas density (thick solid line), pristine gas density
(long-dashed line), AGB ejecta density (thin solid line), SG stellar density
(dashed line).
}
\label{fig:cavcomp} 
\end{figure}

Additional interesting details of this model are illustrated in
Fig. \ref{fig:cavcomp}, where the radial profiles of several
quantities are displayed at two different times, namely $t=40$ Myr and
$t=45$ Myr.  After 40 Myr the pristine gas has already largely moved
toward the GC center.  However, the GC central volume included within
$r\leq 10$ pc still hosts only gas ejected by AGB stars, and the SG
stars formed in this volume till this time have thus the same chemical
characteristics of these polluters. Successively, up to $t\sim 50$ Myr
(see Fig. \ref{fig:cav}), the pristine gas occupies all the GC volume
and is well mixed with the gas returned by the FG stars. The stars
forming in this period suffer a ``diluted'' pollution. Finally, for
$t>50$ Myr, the FG stellar ejecta are again the only source of gas for
the SG star formation and the chemical composition of the new SG stars
formed after this time is that of the FG progenitors' ejecta.

At $t=40$ Myr the SG stars formed at $r<10$ pc have a mass $\sim 8.5
\times 10^4 \msun$, and they are polluted by AGB stars with
masses $M\geq 7$ $\msun$ which are strong He polluters. For $40<t<50$
Myr the chemical contamination of the SG stars is diluted by the
pristine gas all over the GC, and is due to FG stars with masses of
6--7 $\msun$.

\subsection{Implications of the pristine gas infall for the very helium-rich
  population}

Although the study of models including the infall of pristine gas will
need to be extended and further explored, a number of interesting
results have already emerged from the preliminary simulation
discussed in the previous section.

We have shown that there is an epoch, between the end of
SN II phase and the time when the pristine gas begins to mix
with the super-AGB gas ejecta, during which, if the
cluster is massive enough, the winds from the super-AGB stars are
the only contributors to the cooling flow and the SG formation. We
suggest that this is the phase during which the most extreme very
helium-rich SG population can be formed. The super-AGBs can indeed have 
Y$\sim$0.38 in their ejecta, and their gas will not be diluted.  Thus
a very homogeneous population with very high helium will be formed.

This early phase of very helium-rich SG star formation is followed by
an epoch during which  the pristine gas reaches the cluster inner
regions and the  SG
stars form from gas in which the helium of the AGB ejecta, collected
in the inner regions by the cooling flow, 
is diluted with the primordial helium. If the cooling flow goes on
after the pristine gas is exhausted, stars will again be formed with
undiluted ejecta, but the evolving AGB mass and its Y will be smaller
\citep[e.g.][]{pumo2008}, giving rise to a ``helium gap'' between
the first very high helium stars and the other SG stars.

The scenario outlined here might lead to the formation of the blue
main sequences in \ocen\ and in NGC~2808, and of the extremely blue HB
stars in other clusters \citep[for a parametric approach to this
  problem, see][]{bekki2007}.  
\begin{figure}
\centering{
\includegraphics[width=8cm]{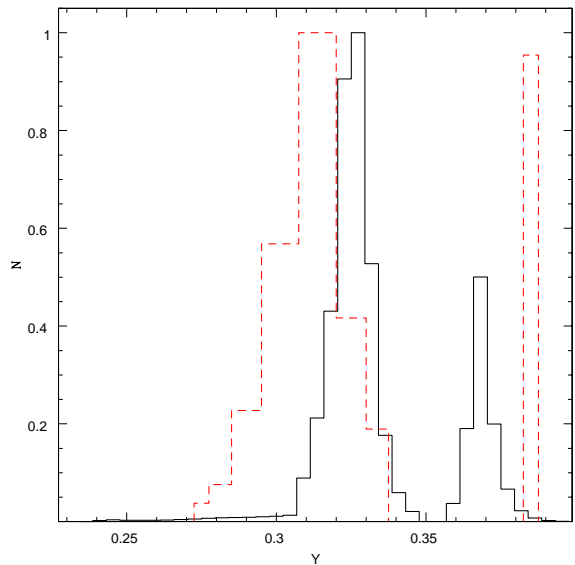}
}
\caption{Histogram of the helium abundance of the SG stars formed in the
  model with the pristine gas infall (solid line). The dashed line
  shows the histogram of Y for NGC~2808 inferred by D'Antona and Caloi
  (2008) from the 
  observations of the HB by Bedin et al. (2004) and from the MS
  distribution by Piotto et al. (2007). Each histogram is normalized
  to its maximum value.}
\label{figndielio}
\end{figure}
 
We implemented in the code the helium abundance of the ejecta as a
function of the initial stellar mass, following the \citet{pumo2008}
compilation, based on the AGB models by \cite{ventura2008a} and on the
super-AGB models by \cite{siess2007}. Fig. \ref{figndielio} shows the predicted 
SG helium distribution for the infall model. 

We see that a peak in the mass distribution
is obtained at Y$\sim 0.37$, corresponding to the star formation in
the cooling flow preceding the infall of pristine matter, and another,
well distinct broader peak at medium--He (Y$\sim 0.31-0.34$) is
obtained from star formation when the wind matter is mixed with the
pristine matter.  There is a negligible tail of stars formed at
Y$<0.25$, so that this model implies that the 50\% stars
having normal helium in NGC~2808 must indeed be a remnant of the FG
population. 
In Fig. \ref{figndielio} we also show the Y distribution derived
for the SG of the cluster NGC~2808 by \cite{dantonacaloi2008}. 

Although we have not made any attempt to find the initial parameters leading to
the best fit of the observational Y distribution, the theoretical and
the observed distribution are, in general, remarkably similar.

Since the helium yields from super-AGBs and AGBs are still uncertain,
the amount of pristine matter involved, and its mixing with the FG
matter is yet to be fully explored, we believe our results show that
this scenario is a very promising step towards the understanding of
the origin of the super-He rich population.

\section{Stellar dynamical evolution}
\label{sec:dyn0}
One of the key requirements of the standard scenario presented in the
previous sections is the large initial mass of the FG population. This
assumption, as we have discussed, is necessary for  the formation of a
significant SG population from the ejecta of intermediate-mass stars. 
The next issue to
address in order to test the viability of the proposed scenario
concerns the possible mechanisms driving the evolution of the ratio of
the number of SG to FG stars.

Numerous studies of the dynamical evolution of star clusters
\citep[see e.g.][and references therein for a review]{heghu03} have
shown that internal two-body relaxation and tidal shocks cause stars
to escape from clusters, eventually leading to complete cluster
dissolution.  With the exception of systems lying within a relatively
narrow range of orbital parameters \citep[see e.g.][]{vh97}, two-body
relaxation is the dominant process driving the evaporation of stars
from clusters.  However the two-body relaxation timescale,
$T_{\rm relax}$, increases with both the mass ($M$) and the size ($R$) of
the system, $T_{\rm relax} \propto M^{1/2} R^{3/2}$ \citep[see
  e.g.][]{heghu03}.  For a cluster with an initial mass of $~10^7
M_{\odot}$, as adopted in the simulations presented in Sect. \ref{sec:res}, the
mass loss induced by internal two-body relaxation in one Hubble time
is negligible.  For example, the dissolution time predicted by
$N$-body simulations \citep{bm03} for a tidally limited
cluster with initial mass $M=10^7 M_{\odot}$, moving on a circular
orbit at 4 kpc from the Galactic center, is about $480$
Gyr.  Hence two-body relaxation cannot be the process responsible for
the loss a significant fraction of the initial FG population, or for
the evolution of the ratio of the number of SG to FG stars.

A more promising mass reduction mechanism for the massive stellar
systems considered in this study is the early mass loss associated
with SN explosions in the FG population.  In response to the loss of
SN ejecta, a stellar system can expand beyond its tidal
limit \citep{cw90, fuhe95}, losing stars from its outer layers. The
fraction of mass lost due to this expansion, and the implications for
a cluster evolution, depend on the cluster structure, IMF and initial
degree of mass segregation.

For initially unsegregated, tidally limited clusters with
a Salpeter IMF, it has been shown that the mass loss resulting from
this mechanism  is negligible if the clusters are sufficiently
centrally concentrated, with concentrations (as measured by the
concentration parameter $c$ defined earlier) similar to those typical
($1\simlt c \simlt 2$; \cite{mcvdm05})   
of Galactic globular clusters \citep{vh97,mcmpz03, bm03}.
However, low-concentration clusters can lose a significant
fraction of their mass in this way, and quickly
dissolve \citep[][see also \cite{vz03} for the possible
  implications of this dissolution on populations of globular
  clusters.]{cs87, cw90, fuhe95}

Both the response of a cluster to this early mass loss and the
critical concentration for its survival depend sensitively on the
initial spatial distribution of massive stars.  Several observational
and theoretical studies \citep[e.g.][]{hille97, hille98, fisch98,
  deg02, sirianni02, gou04, stolte06, bon01, mvpz07} have suggested
that clusters may form with a significant degree of initial mass
segregation imprinted by the star formation process, or created
dynamically very early in their evolution. As discussed in
\cite{vmpz08a, vmpz08b}, initial mass segregation has dramatic
implications for early cluster evolution.  Specifically, in an
initially segregated cluster, the mass lost due to SN explosions is
preferentially removed from the innermost regions of the cluster, and
can lead to rapid and significant overall expansion, mass loss, and
possibly dissolution, even for an initially highly concentrated
system.  Our N-body simulations do not include primordial gas; the
effect of primordial gas expulsion would add to that of SN ejecta in
triggering the cluster early expansion \citep[see e.g.][]{boily2003,bk2007}.

In the following sections we explore the role of this early expansion
on the evolution of multiple population clusters and, in particular,
its implications for the evolution of the cluster stellar populations
relative abundance.

We begin with a general qualitative discussion of the scenario we have
explored then illustrate the process using the results $N$-body
simulations.

\subsection{Cluster mass loss}
\label{sec:dyn1}

In this section we outline the main elements of the dynamical mass
loss scenario, the processes and timescales involved, and
the effect on the numbers of stars in the two populations.

The early evolution of the cluster is characterized by the rapid loss
of FG stars.  This is a consequence of cluster expansion in response
to the dynamical heating from the loss of SN ejecta from the FG
population: as the cluster expands beyond its tidal radius, the outer
layers, which are predominantly populated by FG stars, are
stripped. SG stars, which are mostly confined to the central regions,
remain approximately constant in number during this phase. As a
result, the number ratio of SG to FG stars increases rapidly.
Hereafter, we will refer to the number ratio of long-lived SG to FG
main-sequence stars with masses in the range $0.1<M/M_{\odot}<0.8$ as
$\popms$.

The characteristic timescale for this early evolution, expansion and
stripping is the cluster dynamical time, $T_{dyn} \propto
M^{-1/2}R^{3/2}$, although the precise duration of this phase may
range from $ \sim 10$ to $\sim 10^2$ dynamical times depending on the
degree of initial mass segregation and the amount of impulsive mass
loss; see e.g. Fig. 3 in \cite{vmpz08a}.

The strength of this early expansion and the extent of the consequent
loss of FG stars depend mainly on the degree of initial mass
segregation, the amount of ¬impulsive mass loss due SN and the
structure of the cluster. As mentioned above, for clusters
initially following a King density profile \citep[][]{king66} with
concentrations similar  
to those currently observed in the Galactic globular cluster system
initial mass 
segregation plays a key role in triggering a strong expansion and
early mass loss. As shown by \cite{vmpz08a, vmpz08b} the heating due
to a given amount of impulsive loss of SN ejecta is augmented by the
preferential removal of the mass from the central regions.

A second important initial parameter is the ratio of the cluster ``King
radius'', $r_K$--that is the limiting radius of the King model used to
describe the sytem-- 
to the cluster's tidal radius, $r_{\rm tidal}$ in the Galactic field:
if $r_K/r_{\rm tidal}<1$, the 
cluster  will not lose a significant number of stars until it 
fills its Roche lobe. For a cluster initially undefilling their Roche
lobe, $\popms$ will start to evolve 
later and will
eventually level off at a lower value than would be the case for a
cluster with the same degree
of initial mass segregation but initially filling the
Roche lobe.

As the early expansion phase slows and eventually stops so do the loss
rate of FG stars and the time evolution of $\popms$.  The subsequent
``late'' evolution is driven by two-body relaxation and proceeds on a
timescale, $T_{relax}$, which is significantly longer than the
dynamical timescale. Two-body relaxation is responsible for mixing the
SG and FG populations and for additional mass loss.

For the massive clusters of principal interest in this paper, the
total mass loss due to two-body relaxation is small; for example,
using the mass loss rates derived from N-body simulations by
\cite{bm03} and the current observational values of mass and
galactocentric distance, we estimate that a cluster like \ocen should
have lost at most about 5--10 percent of its initial mass in one
Hubble time by two-body relaxation.

The study of less massive multiple population clusters undergoing a
significant relaxation driven mass loss will be the subject of a
future paper.  We note, however, that star loss induced by two-body
relaxation affects all cluster regions and not just the stars in the
outer layers as in the early phases of the cluster evolution. Both SG
and FG stars escape as a result of two-body relaxation with the result
that the variation of the relative abundance of the two populations is
significantly smaller during this late relaxation-driven phase than
during the early dynamical cluster expansion.

Although relaxation-driven stellar escape has a smaller effect on
the relative abundance $\popms$ of the two populations, the evolution
of this ratio may be affected by mixing, another process driven by
two-body relaxation. The relevant timescale in this case is the
two-body relaxation time of the SG subsystem.  As the low-mass SG
stars evolve toward energy equipartition, the SG subsystem tends to
expand, populating the outer regions of the cluster and mixing with FG
stars.  Once the two populations are mixed, even if the cluster is
still undergoing a strong mass loss, the escaper population is no
longer preferentially composed of FG stars and the number ratio of SG
to FG stars stops increasing.

\subsection{N-body simulations: initial conditions}
\label{sec:dyn2}
We now present the results of $N$-body simulations aimed at
quantitatively exploring the scenario outlined above.

We start our simulations with an initially segregated FG cluster of
25k stars. Mass segregation is set up by first letting the cluster
evolve without including the effects of stellar evolution until the
desired degree of segregation was reached by normal two-body
relaxation.  We emphasize that this is just the procedure used to
generate a self-consistent initially segregated cluster model. It is
not intended to model any stage of cluster evolution.  The degree of
initial segregation is such that the ratio of the initial half-mass
radius of the entire system ($r_h$) to the half-mass radius of stars
more massive than $1~M_{\odot}$ ($r_{h,1}$) is $r_h/r_{h1} \simeq
1.5$.

\begin{figure}
\centering{
\includegraphics[width=8cm]{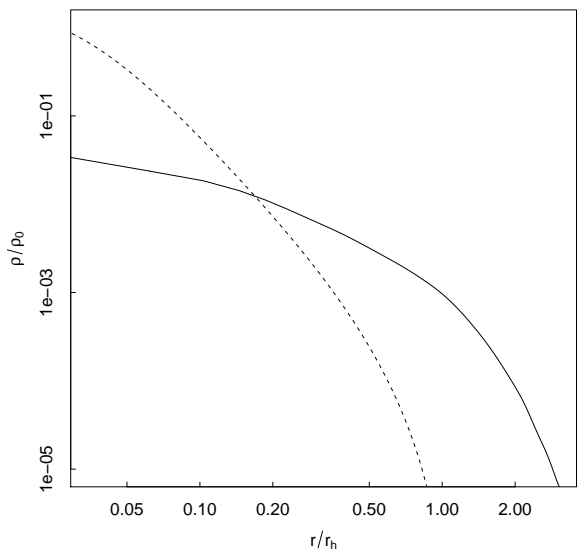}
}
\caption{Initial density (normalized to the total central density
  $\rho_0$) profiles of FG (solid line) and SG (dashed line) stars.}  
\label{densprof}
\end{figure}

Our model cluster is placed on a circular orbit at a galactocentric
distance of $R_g=4$ kpc, with the Galactic tidal field modeled as a
simple Keplerian potential of a point mass $M_g\sim 4.3\times 10^{10}
M_{\odot}$ (corresponding to a circular velocity $v_g\sim 220$ km/s).
We consider both a Roche-lobe-filing system ($r_{\rm K}=r_{\rm
  tidal}$), and three cases in which the system initially underfills
its Roche lobe: $r_{\rm K}/r_{\rm tidal}=0.75$, $r_{\rm K}/r_{\rm
  tidal}=0.6$ and $r_{\rm K}/r_{\rm tidal}=0.5$.  Hereafter we refer
to these three simulations as \sgnAA, \sgnBB, \sgnCC, and \sgnD
respectively.

SG stars are inserted into the simulation in a simplified way, by
replacing, after about 30 Myr, the progenitor population of FG
intermediate-mass stars ($4<M/M_{\odot}<8$) by a SG population having
masses between $0.1~M_{\odot}$ and $0.8~M_{\odot}$ drawn from a Kroupa
IMF.  Following the results of the hydrodynamical simulations
presented in Sect. \ref{sec:res}, newborn SG stars are concentrated in
the innermost regions of the cluster. In the $N$-body simulations, the
SG stars are arbitrarily distributed according to a high-concentration
$W_0=7$ ($c=1.5$) King model, with a half-mass radius one-tenth of
that of the initial FG system.  The SG population is introduced with
negligible velocity dispersion, and so undergoes an initial rapid
phase of violent relaxation.  Largely for this reason, the details of
initial SG density distribution are unimportant in determining the
subsequent evolution of the SG/FG system.  The adopted initial density
profiles of the two stellar populations are shown in
Fig.~\ref{densprof}.

As shown by analytical calculations and $N$-body simulations presented
in \cite{vmpz08a, vmpz08b}, the heating due to mass loss from (FG) SN
ejecta increases both with the degree of initial mass segregation of
the objects losing the mass and with the amount of impulsive mass
loss.  In order to clarify the implications of a lower level of
initial mass segregation of the FG population and/or of a smaller
amount of early impulsive mass loss from FG SN ejecta, we have
repeated the \sgnA simulation after decreasing the velocities of all
FG stars so to reduce the initial virial ratio by approximately 10 per
cent and 30 per cent (changing directly the initial virial ratio of
the system allows us to easily focus on and explore the effect of a
different initial degree of mass segregation while leaving all the
other cluster properties unchanged).  Hereafter, we refer to these two
simulations as \sgnco and \sgnccoA.

As described in the previous section, the scenario we are exploring is
one in which cluster evolution is driven by multiple physical
processes operating on different timescales. For the massive systems
of interest here, all the relevant timescales (the dynamical time, the SG
subsystem relaxation time and the cluster relaxation time) are
significantly different from one another. In particular, the early
dynamical evolution phase of the cluster is quite distinct from the
later phase of relaxation-driven mixing and mass loss. In order to
maintain this clear separation between the early and late
evolutionary phases for the relatively small-$N$ system we can
actually study, we have 
introduced into our $N$-body simulations a softening parameter equal to the
average interparticle distance within the cluster half-mass radius  in
the calculation of the   
force. The effect of this softening is to increase
the relaxation timescale, avoiding the overlap between  early rapid dynamical
processes and late mixing and mass loss driven by two-body relaxation
that would otherwise be found in the $N$-body models.

All simulations were carried out using the {\tt Starlab} package
\citep[{\tt http://www.manybody.org/}]{pz01} accelerated by GRAPE-6
special-purpose hardware \citep{makino03}. {\tt Starlab} includes the
effects of mass loss due to stellar evolution, two-body relaxation,
and the influence of the Galactic tidal field.

\subsection{N-body simulations: results} 
\label{sec:dyn3}

\begin{figure}
\centering{
\includegraphics[width=8cm]{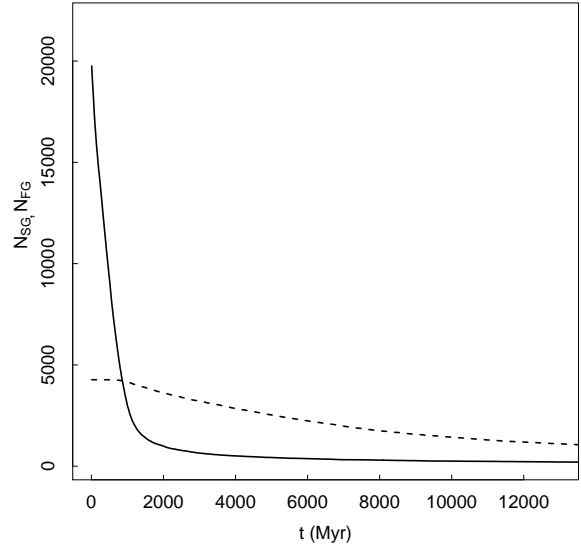}
}
\caption{Time evolution of the number of SG (dashed line), FG
  (solid line)  main sequence stars with $0.1<M/M_{\odot}<0.8$ for the \sgnA simulation.} 
\label{numberevol}
\end{figure}

Fig.~\ref{numberevol} shows
the time evolution of the number of SG, $N_{SG,ms}$, and FG, $N_{FG,ms}$
main sequence stars with 
$0.1<M/M_{\odot}<0.8$ for the \sgnA  simulation. The rapid preferential
loss of FG stars during the cluster early evolution, discussed in
Sect. \ref{sec:dyn1}, is clearly evident. As the cluster expands beyond its
tidal radius in response
to the loss of FG SN ejecta, its outer layers are stripped and a large
fraction of the initial FG population is lost.
The SG subsystem, on the other hand, is concentrated in the cluster
inner regions and is largely unscathed by this early evolutionary
process. 

\begin{figure}
\centering{
\includegraphics[width=8cm]{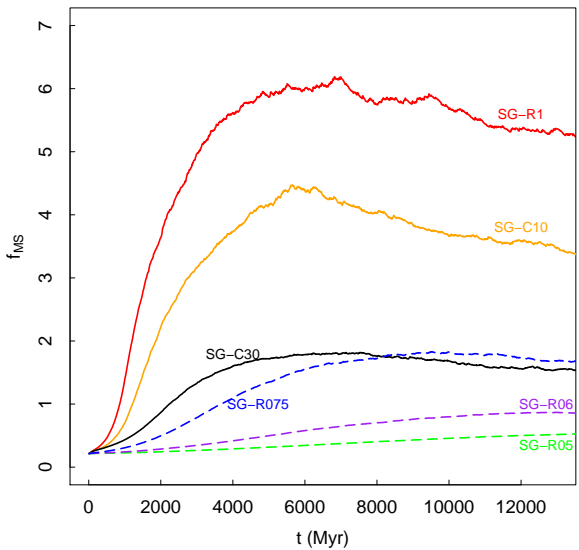}
}
\caption{ Time evolution of $\popms$, the ratio of the number of SG to FG  main
sequence stars with $0.1<M/M_{\odot}<0.8$, for all the
$N$-body simulations presented in the paper.
}
\label{ratioevol}
\end{figure}

The early cluster expansion and associated mass loss slow
significantly as the cluster gradually enters the phase driven by
two-body relaxation.  The time evolution of $\popms =
N_{SG,ms}/N_{FG,ms}$ for all the models explored in our simulations is
shown in Fig.~\ref{ratioevol}.  The early strong preferential loss of
FG stars leads to a rapid increase of $\popms$; subsequently when the
cluster expansion and mass loss slow down, the growth rate of $\popms$
gradually decreases.  Fig.~\ref{ratioevol} clearly illustrates the
role of initial conditions in determining the current value of
$\popms$ and suggests a possible origin for cluster-to-cluster
variation in the relative abundance of different populations.  Our
simulations indicate that the typical observed values of $\popms \sim
0.5-1.5$ are attained by clusters initially underfilling their Roche
lobe and/or with moderate levels of initial mass segregation.

Interestingly, our simulations also show  possible
evolutionary routes to the loss of most of the FG
populations, leaving a cluster with a SG-dominated population. 
Initial conditions triggering the strong expansion  
necessary to strip most of the FG population  are
characterized by a larger degree of initial mass segregation and/or a
larger amount of impulsive mass loss. The amount of
impulsive mass loss will differ, for example, in clusters with the same 
stellar IMF and the same level of mass segregation but with different
initial sizes (for example tidally truncated clusters at different
galactocentric distances) and therefore different dynamical
times. These clusters 
would lose the same amount of FG mass due to SN ejecta but for clusters
with longer dynamical times, a larger fraction of this mass loss
would occur in the impulsive regime, contributing to the heating that
drives the cluster initial expansion. 

As the structure and the abundances of the FG and SG populations
evolve, the relative spatial distribution of the two
populations also undergo a significant variation. 

\begin{figure}
\centering{
\includegraphics[width=8cm]{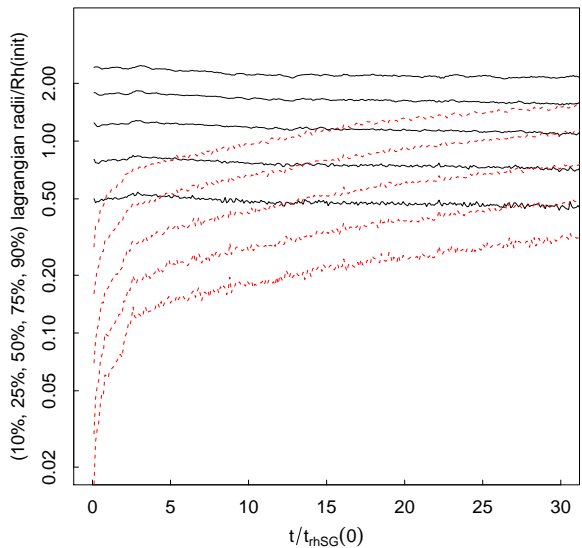}
}
\caption{Time evolution  of the 10, 25, 50, 75, 90 per cent lagrangian
radii of the FG (solid line) and the SG (dashed line) population of
main sequence stars with $0.1<M/M_{\odot}<0.8$ for the \sgnA simulation. $t_{rh,SG}(0)$ is the initial half-mass relaxation time of the SG subsystem.}
\label{lagrevol}
\end{figure}

Fig.~\ref{lagrevol} shows the evolution of the 10, 25, 50, 75 and 90
per cent lagrangian radii for FG and SG main sequence stars with
masses in the range $0.1<M/M_{\odot}< 0.8$ for the \sgnA simulation
(repeated in this case without softening in order to properly
interpret the evolutionary timescale in terms of the SG subsystem
initial half-mass relaxation time).

\begin{figure}
\centering{
\includegraphics[width=8cm]{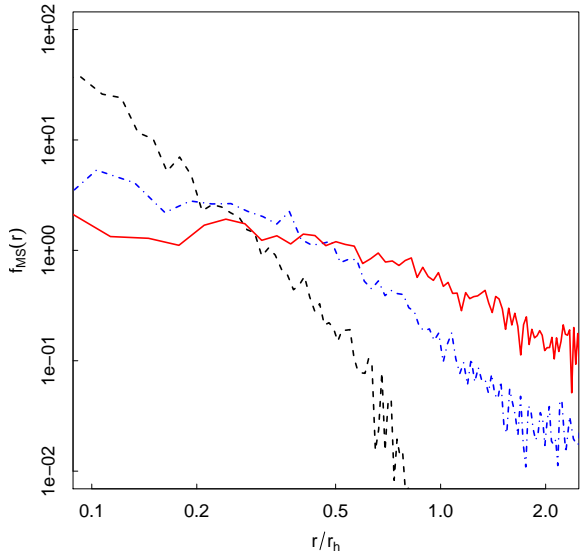}
}
\caption{Radial profile of $\popms$ at $t/t_{rh,SG}(0)=0, 10, 25$ (dashed black
  line, dot-dashed blue line, solid red line) for the \sgnA
  simulation. }
\label{ratioprofile}
\end{figure}

Fig.~\ref{ratioprofile} shows the radial profile of $\popms$ at three
different times during the \sgnA simulation.  Our simulations predict
that an initially strongly centrally peaked $\popms(r)$ profile will
evolve toward a profile that is flat in the inner regions and
decreasing in the outer parts.  As the system evolves and the two
populations continue to mix, the flat part of the profile (i.e. the
region where the two populations have similar number density profiles)
expands.

For a  detailed comparison between the level of mixing in real clusters and that
predicted by our simulations, we need an estimate of the ratio of the
cluster age, $T$, to the 
initial SG subsystem half-mass relaxation time, $t_{rh,SG}(0)$. 
The lack
of strong constraints on the initial conditions complicates such a
comparison. However if we plausibly adopt the current central
relaxation time of \ocen  \citep[$\sim 1$ Gyr, see e.g.][]{har96} as  an
estimate of the initial half-mass relaxation time of the SG
subsystem,  $T/t_{rh,SG}(0) \approx 10$. The radial
profile of $\popms$, predicted by our simulations for this value of
$T/t_{rh,SG}(0)$ is shown in Fig.\ref{ratioprofile}. It is
characterized by a central flat region where SG
and FG stars are mixed, and declines in the
outer cluster regions still dominated by  FG main sequence stars.  
 
We note that the presence of a declining outer part in the profile
of $\popms(r)$ is 
qualitatively consistent with the observed radial profile 
of the number ratio of blue to red main sequence stars in \ocen
\citep[][]{sollima07}. 
Observations of the
relative abundance  of different populations in the
inner regions of \ocen and other  multiple population clusters are needed to
test the inner flattening predicted by our simulations. 
It is important to point out that while our predictions concerning the shape and
evolution  of the $\popms(r)$ profile are general, the 
 quantitative characteristics of this profile (e.g.
the slope of the outer parts and the extension of the flat part of the
profile)  depend on  
several parameters including the 
initial relative concentration of the two populations, the details
of their initial density profiles, the initial
degree of mass segregation, the number of stars left in the cluster
after the early mass loss and the cluster relaxation time. We therefore
emphasize that a quantitative comparison between our theoretical
$\popms(r)$ profiles and  
observations must await the results of a more extended survey of
simulations further exploring the space of initial parameters.

Our simulations show that both $\popms$ and the relative concentration
of the two populations are significantly affected by dynamical
evolution, and both depend on the cluster dynamical history and its
initial conditions.  We emphasize that the goal of the simulations
reported here is to illustrate the key ingredients and dynamical
processes behind the evolution of the relative number and spatial
distribution of FG and SG stars.  A more thorough survey of the
parameters relevant to the scenario presented here is needed for a
detailed comparison with observational data, and is currently in
progress.  The results will be presented elsewhere.

\section{Conclusions}
In this paper we have explored a scenario for the formation and the
dynamical evolution of multiple stellar generations in globular
clusters.  A successful model for the origin and the evolution of
clusters with multiple stellar generations needs to satisfy a number
of observational constraints concerning the current chemical
properties, relative abundance and structural properties of the
different stellar populations.

After reviewing the main observational results on multiple population
clusters, we have pointed out that the key issues to address concern
the general origin of SG stars (among which those with very high
helium content observed in the most massive clusters are only a
fraction) and the identification of the evolutionary processes leading
to clusters with a similar number of SG and FG stars (or even to
clusters dominated by SG stars).

In order to address these different constraints and be able to model
all the relevant aspects of multiple population clusters, we have
combined the input from stellar evolution models to constrain the
origin  of the gas from which SG stars might form, hydrodynamical
simulations to model the process of SG star formation and predict its
initial structural properties and, finally, $N$-body simulations to
explore the subsequent dynamical evolution of the two populations.

The main steps and results of this investigation are as follows.

\subsection*{1.~Chemical Abundances}
We have reviewed all the issues concerning the super-AGB and massive
AGB stars as the source of the polluted gas from which SG stars could
form.  The hypothesis that AGB stars might provide the gas for the SG
formation has recently received some criticism in the literature based
on possible inconsistencies between models and the observed chemical
abundances in SG stars.  However, in our review, we have pointed out
that the results of chemical processing during the Hot Bottom Burning
phases are highly affected by the uncertainties in the basic input
physics of these models, and that there is an ample range of plausible
input parameters which can provide yields consistent with
observations. In addition, recent models show that the super-AGB
evolution can produce the helium content required for the very
helium-rich population mentioned above. During the super-AGB epoch,
between the end of SN II explosions and the beginning of the SN Ia
explosions, there are no mechanisms preventing the formation of a
cooling flow and the formations of the very helium-rich SG population
from super-AGB winds.

\subsection*{2.~Hydrodynamical Simulations}
We have carried out a number of hydrodynamical simulations to explore
the dynamics of gas from super-AGB and AGB stellar winds and the
process of formation of the SG population in the assumption that the
SNe II belonging to the FG population clear the GC of its pristine
gas.

\begin{enumerate}

\item Our simulations show that this gas collects in a cooling flow
  into the innermost regions of the cluster where it forms SG
  stars. The cluster emerging from the hydrodynamical simulations is
  one with a SG strongly concentrated in the inner core of a more
  extended FG population. Initially the FG stars are the dominant
  stellar population with a total mass that must be about ten times
  larger than the total mass of the SG stars. The initial mass of the
  FG stars needs to be large enough to provide enough stellar mass
  return and form a substantial amount of SG stars.

\item We have explored the dependence of our results on the initial
  mass of the cluster, the stellar IMF and the initial cluster size
  and concentration.  SG stars are always concentrated in the cluster
  innermost regions and the structure of the newly formed SG subsystem
  is largely independent of the cluster size and concentration.

\item Starting with a initial FG mass ten times smaller than the value
  adopted in our standard model, leads to a SG total mass also ten
  times smaller but does not alter the shape of the SG density profile
  or its concentration.  For simulations with a flatter IMF the larger
  amount of gas supplied by AGB winds implies that a smaller star
  formation efficiency is needed to form the required mass of SG
  stars.  If the requirement that the current mass of long-lived SG
  and FG stars are similar is maintained, a large FG initial mass is
  needed also for a flatter IMF. In this case the number of long-lived
  FG stars is smaller, and a large value of $\mi$ is still needed to
  have a current mass of long-lived FG stars consistent with
  observations.
  
\item The effects of SNe Ia and additional energy sources on the SG
  formation process have been investigated. Our simulations show that,
  for a ``normal'' (e.g. Kroupa) IMF of the FG stars, the stellar
  explosions quickly evacuate most of the gas returned by the AGB
  stars, leading to a substantial drop in the gas content of the GC
  and to a suspension of the SG star formation.  Given the
  uncertainties on the SN Ia rate, the interruption of the SF may
  occur in the time interval $40<t<100$ Myr. Flat IMFs produce an
  increment in the amount of returned gas, leading to an enhancement
  of the radiative losses and, possibly, to the inability of the SNe
  Ia to get rid of the gas. In this case SG stars continue to form and
  their chemical abundances should show the signature of the pollution
  by gas from SN ejecta.  This however does not seem to be confirmed
  observationally and flat IMFs are thus disfavoured.

\item We have further considered the possible presence of a diffuse
  energy source due to X-ray binaries and/or planetary nebulae. We
  show that for the standard model the cooling flow and the associated
  SG star formation can be halted by energy sources with a total power
  larger than $10^{38}$ erg s$^{-1}$.

\item We have also explored the possibility that the FG SNe II do not
  vent away all the pristine gas, and that part of it remains
  gravitationally bound to the cluster and falls back onto it after an
  initial ousting.  The SG stars forming in the cluster central
  regions before this pristine gas reaches the cluster core originate
  from pure ejecta of the super-AGB stars which is very helium
  rich. These stars may form the very helium-rich population present
  in the most massive GCs. Once the pristine gas has fallen back in
  the cluster, SG stars form in an ambient medium in which the helium
  abundance is ``diluted,'' and give rise to the moderately helium
  rich population.

\end{enumerate}

\subsection*{3.~Stellar Dynamical Simulations}
The subsequent dynamical evolution of the cluster has been studied by
means of $N$-body simulations.
 
\begin{enumerate}

\item One of the critical issues for viability of the scenario
  presented here is the identification of a mechanism leading to the
  escape of a large fraction of the initial FG population.  For
  clusters as massive as required by our hydrodynamical models, the
  mass lost in one Hubble time due to two-body relaxation is
  negligible.

  Following observational and theoretical studies suggesting that
  initial mass segregation may be imprinted in clusters by the star
  formation process or produced dynamically early in the cluster
  evolution, we have studied the evolution of initially segregated
  clusters.  In our simulations, a large fraction of FG cluster stars
  are lost early in the cluster evolution due to the expansion and
  stripping of the cluster outer layers resulting from early mass loss
  associated with FG SN ejecta. Initial FG mass segregation plays a
  key role since the heating and the expansion due to the loss of SN
  ejecta are augmented by the preferential removal of the mass from
  the inner regions of the cluster.
  
\item The population of escapers is dominated by FG stars. The SG
  population, initially concentrated in the innermost cluster regions,
  is largely unscathed by the early mass loss.  We find that the early
  cluster evolution and mass loss can lead to a significant loss of FG
  stars and to current values of $\popms$, consistent with
  observations ($0.5<\popms<1.5$).  We have also demonstrated possible
  evolutionary routes leading to the loss of most of the FG population
  and leaving a SG-dominated cluster. Clusters initially filling their
  Roche lobe with higher degrees of initial mass segregation and/or
  losing impulsively a larger fraction of SN ejecta undergo more
  expansion and may lose most of their FG stars.

\item After the initial phase of expansion and mass loss, cluster
  evolution is driven by two-body relaxation which results in the
  spatial mixing of the two populations.  Following the hydrodynamical
  simulations, in the $N$-body initial conditions SG stars are
  concentrated in the inner regions of the cluster.  The radial
  profile of the number ratio of SG to FG stars, $f_{MS}(r)$ is
  initially a decreasing function of radius.  We find that, as the
  cluster evolves and the two populations mix, $f_{MS}(r)$ flattens in
  the inner parts of the cluster. The central flat region of the
  $f_{MS}(r)$ radial profile expands on a relaxation timescale as
  mixing proceeds.  In general, unless mixing is complete, we predict
  that $\popms(r)$ is characterized by a flat inner part and a
  declining portion in the outer regions of the cluster. 
\end{enumerate}

\par\smallskip\noindent
A more extended survey of hydrodynamical and $N$-body simulations to
further exlore the dynamics and the implications of the scenario
presented here as well as for a more detailed comparison with
observations is currently in progress and will be presented in future
papers.

\section*{Acknowledgments}
We thank Luca Ciotti for useful discussions.  AD, FD and SR were
supported in part by INAF under PRIN 2005 ``Experimenting stellar
nucleosynthesis in clean environments''.  EV and SM were supported in
part by NASA grants NNX07AG95G and NNX08AH15G, and by NSF grant
AST-0708299.


\section*{Appendix}
\label{sec:app}
The result of our simulations proved to be essentially independent of
the value of $\nu$. To understand why, let us consider a one-zone toy
model in which the gas density $\rho$ and the stellar density $\rho_*$
evolve following the two equations
\begin{equation}
\label{eq:gas}
{d\rho \over dt}=-\nu \rho^2+s
\end{equation}
\begin{equation}
\label{eq:str}
{d\rho_* \over dt}=\nu \rho^2
\end{equation}
The constant $\nu$ regulates the SF efficiency and the constant $s$ is
a source term representing the rate at which the gas is
replenished. The SF timescale $t_{\rm sf}=1/\nu\rho$ depends on the
gas density as the cooling time rather than the free fall time, which
instead is the actual timescale encountered in our simulations. With
the present definition, however, it is possible to obtain an
analytical solution to the above equations without invalidating our
main conclusions (see Sect. \ref{subsec:model}).  The solution of
Eq. \ref{eq:gas} is
\begin{equation}
\label{eq:sgas}
\rho(t)=\left( {s\over \nu}\right )^{0.5}\tanh\left({t\over\tau}\right),
\end{equation}
where $\tau=(\nu s)^{-0.5}$. Then the solution of Eq. \ref{eq:str} is
\begin{equation}
\label{eq:sstr}
\rho_*(t)=\left( {s\over \nu}\right )^{0.5}\left[\left({t\over\tau}\right)-
\tanh\left({t\over\tau}\right)\right].
\end{equation}
Defining the replenishing timescale as $t_{\rm rp}=\rho/s$, one
obtains $\tau=(t_{\rm sf}t_{\rm rp})^{0.5}$. In words, the growth
timescale of the stars is given by the geometric mean of the SF and
replenishing timescales, and is therefore of the same order of
magnitude. This timescale is necessarily much shorter than the
evolution time of the system over which eqs. \ref{eq:gas} and
\ref{eq:str} are solved. In fact, these equations are meaningless for
$t/\tau<1$ because neither appreciable SF nor significant stellar mass
loss can occur during a time interval that short. In the meaningful
case in which $t/\tau\gg 1$, Eq. \ref{eq:sstr} becomes
\begin{equation}
\label{eq:asin}
\rho_*(t)=st.
\end{equation}
The growth of the stellar population is thus independent of $\nu$ and
depends on the replenishing rate, a result confirmed by our
simulations.

\label{lastpage}
\end{document}